\documentclass[preprints,article,accept,moreauthors,pdftex]{mdpi}

\firstpage{1} 
\pubvolume{1}
\issuenum{1}
\articlenumber{0}
\pubyear{2022}
\copyrightyear{2022}
\externaleditor{{Academic Editor: Antonio Ereditato}}  
\datereceived{19 November 2021} 
\dateaccepted{12 January 2022} 
\datepublished{} 
\hreflink{https://doi.org/}

\usepackage{lineno,hyperref}
\usepackage{diagbox}
\usepackage{eurosym}
\usepackage{xspace}
\modulolinenumbers[5]

\newcommand{\fe}{\ensuremath{^{55}\textrm{Fe}}\xspace}

\newcommand{\ambe}{\ensuremath{\textrm{Am} \textrm{Be}}\xspace}

\newcommand{\lemon}{{\textsc{Lemon}}\xspace}
\newcommand{\lime}{{\textsc{Lime}}\xspace}

\newcommand{\gac}{{\textsc{Gac}}\xspace}

\newcommand{\GEANT} {{\textsc{Geant4}}\xspace}
\newcommand{\SRIM} {{\textsc{Srim}}\xspace}

\newcommand{\unit}[1]{\ensuremath{\textrm{\,#1}}\xspace}
\newcommand{\keV}{\ensuremath{\,\textrm{keV}}\xspace}
\newcommand{\eV}{\ensuremath{\,\textrm{eV}}\xspace}
\newcommand{\keVee}{\ensuremath{\,\textrm{keV}_{ee}}\xspace}
\newcommand{\eVee}{\ensuremath{\,\textrm{eV}_{ee}}\xspace}
\newcommand{\keVnr}{\ensuremath{\,\textrm{keV}_{nr}}\xspace}
\newcommand{\keVr}{\ensuremath{\,\textrm{keV}_{r}}\xspace}
\newcommand{\eVnr}{\ensuremath{\,\textrm{eV}_{nr}}\xspace}

\newcommand{\Ed}  {E$_{\mathrm{D}}$\xspace}
\newcommand{\Et}  {E$_{\mathrm{Transf}}$\xspace}
\newcommand{\Vg}  {V$_{\mathrm{GEM}}$\xspace}

\Title{The CYGNO Experiment}

\TitleCitation{The CYGNO Experiment}

\Author{  
\mbox{{Fernando Domingues Amaro}  $^{1}$,}
\mbox{Elisabetta Baracchini $^{2,3}$,}
\mbox{Luigi  Benussi $^{4}$,}
\mbox{Stefano Bianco $^{4}$,}
\mbox{Cesidio Capoccia $^{4}$,}
\mbox{Michele Caponero $^{4,5}$,}
\mbox{Danilo Santos Cardoso $^{6}$,}
\mbox{Gianluca Cavoto $^{7,8}$,}
\mbox{Andr\'e Cortez $^{2,3}$,}
\mbox{Igor Abritta Costa $^{9}$,}
\mbox{Rita Joanna da Cruz Roque $^{1}$,}
\mbox{Emiliano Dan\'e $^{4}$,}
\mbox{Giorgio Dho $^{2,3}$,}
\mbox{Flaminia Di Giambattista $^{2,3}$,}
\mbox{Emanuele Di Marco $^{7}$,}
\mbox{Giovanni Grilli di Cortona $^{4}$,}
\mbox{Giulia D'Imperio $^{7}$,}
\mbox{{Francesco Iacoangeli} $^{7}$,}
\mbox{{Herman Pessoa Lima }J\'unior $^{6}$,}
\mbox{Guilherme Sebastiao Pinheiro Lopes $^{9}$,}
\mbox{Amaro da Silva Lopes Júnior $^{9}$,}
\mbox{{Giovanni Maccarrone} $^{4}$,}
\mbox{Rui Daniel Passos Mano $^{1}$,}
\mbox{{Michela Marafini }$^{10}$,}
\mbox{Robert Renz Marcelo Gregorio $^{11}$,}
\mbox{David Jos\'e Gaspar  Marques $^{2,3}$,}
\mbox{{Giovanni Mazzitelli }$^{4}$,}
\mbox{{Alasdair Gregor McLean} $^{11}$,}
\mbox{{Andrea Messina} $^{7,8}$,}
\mbox{Cristina Maria Bernardes Monteiro $^{1}$,}
\mbox{Rafael Antunes Nobrega $^{9}$,}
\mbox{Igor Fonseca Pains $^{9}$,}
\mbox{Emiliano Paoletti $^{4}$,}
\mbox{Luciano Passamonti $^{4}$,}
\mbox{Sandro Pelosi $^{7}$,}
\mbox{Fabrizio Petrucci $^{12,13}$,}
\mbox{Stefano Piacentini $^{7,8}$,}
\mbox{Davide Piccolo $^{4}$,}
\mbox{Daniele Pierluigi $^{4}$,}
\mbox{Davide Pinci $^{7,}$*,}
\mbox{Atul Prajapati $^{2,3}$,}
\mbox{{Francesco Renga }$^{7}$,}
\mbox{Filippo Rosatelli $^{4}$,}
\mbox{{Alessandro Russo }$^{4}$,}
\mbox{Joaquim Marques Ferreira dos Santos $^{1}$,}
\mbox{{Giovanna Saviano} $^{4,14}$,}
\mbox{Neil John Curwen Spooner $^{11}$,}
\mbox{Roberto Tesauro $^{4}$,}
\mbox{Sandro Tomassini $^{4}$} and~\mbox{{Samuele Torelli} $^{2,3}$}
} 

\AuthorNames{Fernando Domingues Amaro,
Elisabetta Baracchini,
Luigi  Benussi,
Stefano Bianco,
Cesidio Capoccia,
Michele Caponero,
Danilos Santos Cardoso,
Gianluca Cavoto,
Andr\'e Cortez,
Igor Abritta Costa ,
Emiliano Dan\'e,
Rita Joanna da Cruz Roque,
Giorgio Dho,
Flaminia Di Giambattista,
Emanuele Di Marco,
Giulia D'Imperio,
Francesco Iacoangeli,
Herman Pessoa Lima J\'unior,
Guilherme Sebastiao Pinheiro Lopes,
Amaro da Silva Lopes J\'unior ,
Giovanni Maccarrone,
Rui Daniel Passos Mano,
Michela Marafini,
Robert Renz Marcelo Gregorio,
Giovanni Mazzitelli,
Alasdair Gregor McLean,
Andrea Messina,
Cristina Maria Bernardes Monteiro,
Rafael Antunes Nobrega,
Igor Fonseca Pains,
Emiliano Paoletti,
Luciano Passamonti,
Sandro Pelosi,
Fabrizio Petrucci,
Stefano Piacentini,
Davide Piccolo,
Daniele Pierluigi,
Davide Pinci,
Atul Prajapati,
Francesco Renga,
Filippo Rosatelli,
Alessandro Russo,
Joaquim Marques Ferreira dos Santos,
Giovanna Saviano,
Neil John Curwen Spooner,
Roberto Tesauro,
Sandro Tomassini and
Samuele Torelli.}

\AuthorCitation{Amaro, F.D.; Baracchini, E.; Benussi, L.; Bernardes Monteiro, C.M.; Bianco, S.; Capoccia, C.; Caponero, M.; Cardoso D.S.; Cavoto, G.; Cortez, A.;  et al.
}

\address{%
$^{1}$ \quad LIBPhys, Department of Physics, University of Coimbra, 3004-516 Coimbra, Portugal; famaro@uc.pt (F.D.A.); cristinam@uc.pt (C.M.B.M.);  RDPMano@uc.pt (R.D.P.M.);  ritaroque@fis.uc.pt (R.C.R.); \mbox{jmf@uc.pt (J.M.F.d.S.)}\\

$^{2}$ \quad Gran Sasso Science Institute, 67100  L'Aquila, Italy; 
 elisabetta.baracchini@gssi.it (E.B.);  andre.f.cortez@gmail.com (A.C.); \mbox{giorgio.dho@gssi.it (G.D.)}; flaminia.digiambattista@gssi.it (F.D.G.); david.marques@gssi.it (D.J.G.M.); atul.prajapati@gssi.it (A.P.); samuele.torelli@gssi.it (S.T.) \\

$^{3}$ \quad Istituto Nazionale di Fisica Nucleare, Laboratori Nazionali del Gran Sasso, 67100  Assergi, Italy \\

$^{4}$ \quad Istituto Nazionale di Fisica Nucleare, Laboratori Nazionali  di Frascati,  00044, Frascati, Italy; \mbox{luigi.benussi@lnf.infn.it (L.B.)}; stefano.bianco@lnf.infn.it (S.B.); Cesidio.Capoccia@lnf.infn.it (C.C.); 
michele.caponero@enea.it (M.C.); Emiliano.Dane@lnf.infn.it (E.D.); 
grillidc@lnf.infn.it (G.G.d.C.);
giovanni.maccarrone@lnf.infn.it (G.M.); giovanni.mazzitelli@lnf.infn.it (G.M.); Emiliano.Paoletti@lnf.infn.it~(E.P.); \mbox{luciano.passamonti@lnf.infn.it~(L.P.)}; Davide.Piccolo@lnf.infn.it~(D.P.); {Daniele.Pierluigi@lnf.infn.it~}(D.P.); filippo.rosatelli@lnf.infn.it (F.R.); arusso@lnf.infn.it (A.R.); giovanna.saviano@cern.ch (G.S.); Roberto.Tesauro@lnf.infn.it (R.T.); sandro.tomassini@lnf.infn.it (S.T.)\\ 
$^{5}$ \quad ENEA Centro Ricerche Frascati, 00044  Frascati, Italy \\

$^{6}$\quad Centro Brasileiro de Pesquisas Físicas, Rio de Janeiro 22290-180, RJ, Brazil; 
 dsantoscardoso@outlook.com~(D.S.C.); hlima@cbpf.br (H.P.L.J.); \\

$^{7}$ \quad Istituto Nazionale di Fisica Nucleare, Sezione di Roma, 00185  Roma, Italy; 
gianluca.cavoto@roma1.infn.it~(G.C.); emanuele.dimarco@roma1.infn.it~(E.D.M.); giulia.dimperio@roma1.infn.it (G.D.); francesco.iacoangeli@roma1.infn.it (F.I.); andrea.messina@uniroma1.it~(A.M.); Alessandro.Pelosi@roma1.infn.it (S.P.); stefano.piacentini@uniroma1.it (S.P.); francesco.renga@roma1.infn.it (F.R.)\\

$^{8}$ \quad Dipartimento di Fisica, Sapienza Universit\`a di Roma, 00185  Roma, Italy \\

$^{9}$ \quad Universidade Federal de Juiz de Fora, Faculdade de Engenharia, Juiz de Fora 36036-900,   MG, Brazil; igorabritta@gmail.com (I.A.C.); guilherme.lopes@engenharia.ufjf.br (G.S.P.L);  amaro.lopes@engenharia.ufjf.br (A.d.S.L.J.) rafael.nobrega@ufjf.edu.br (R.A.N.);igor.pains@engenharia.ufjf.br (I.F.P.); \\

$^{10}$\quad Museo Storico della Fisica e Centro Studi e Ricerche ``Enrico Fermi'', Piazza del Viminale 1, 00184  Roma, Italy; Michela.Marafini@roma1.infn.it \\

$^{11}$\quad Department of Physics and Astronomy, University of Sheffield, Sheffield S3 7RH, UK; \mbox{robert.gregorio@sheffield.ac.uk (R.R.M.G.)}; ali.mclean@sheffield.ac.uk (A.G.M.); \mbox{n.spooner@sheffield.ac.uk (N.J.C.S.)} \\
$^{12}$\quad Dipartimento di Matematica e Fisica, Universit\`a Roma TRE, 00146  Roma, Italy; fabrizio.petrucci@uniroma3.it   \\
$^{13}$\quad Istituto Nazionale di Fisica Nucleare, Sezione di Roma TRE, 00146  Roma, Italy \\
$^{14}$\quad Dipartimento di Ingegneria Chimica, Materiali e Ambiente, Sapienza Universit\`a di Roma, 00185  Roma, Italy \\
}


\corres{Correspondence: davide.pinci@roma1.infn.it}

\abstract{The search for a novel technology able to detect and reconstruct nuclear and electron recoil events with the energy of a few \keV~has become more and more important now that large regions of high-mass dark matter (DM) candidates have been excluded.
Moreover, a detector sensitive to incoming particle direction will be crucial in the case of DM discovery to open the possibility of studying its properties.
Gaseous time projection chambers (TPC) with optical readout are very promising detectors combining the detailed event information provided by the TPC technique with the high sensitivity and granularity of latest-generation scientific light sensors.
The CYGNO experiment (a CYGNus module with Optical readout) aims to exploit the optical readout approach of multiple-GEM structures in large volume TPCs for the study of rare events as interactions of low-mass DM or solar neutrinos.
The combined use of high-granularity  sCMOS cameras and
fast light sensors allows the reconstruction of the 3D direction of the tracks, offering good energy resolution and very high sensitivity in the few keV energy range, together with a very good particle identification useful for distinguishing nuclear recoils from electronic
recoils.
This experiment is part of the CYGNUS proto-collaboration, which aims at constructing a network of underground observatories for directional DM search. 
A one cubic meter demonstrator is expected to be built in 2022/23 aiming at a larger scale apparatus (30 m$^3$--100 m$^3$) 
at a later stage.}


\begin{document}
\section{Introduction}


The presence in the universe of large amounts of non-luminous matter (usually referred to as dark matter (DM)) is nowadays an established, yet still mysterious, paradigm~\cite{Bertone:2004pz}. Deciphering its essence is one of the most compelling tasks for fundamental physics today. Electrically neutral and very low interacting particles with a mass in the range of few to thousands of GeV are usually referred to as {\it weakly interacting massive particles (WIMPs}), and represent a well-motivated DM candidate independently predicted by the extension of the Standard Model of particle physics and the $\Lambda$-CDM model of cosmology. The~measurements of the rotational curve of our galaxy suggests the presence of a DM halo, through which ordinary, luminous galactic matter is travelling in its rotation around the galactic center. This creates a relative motion between an observer on Earth and the particles in the halo that scientists seek to exploit to detect DM through their elastic scattering with ordinary matter. In~particular,  low-energy (1--100 keV) nuclear recoils (NR) are expected to be the clearest evidence of WIMP~interactions.

Given their rarity, the~main experimental challenge of direct DM searches in the GeV mass region is to discriminate NR from interactions induced by other particles, which have largely higher rates. The~apparent WIMP wind would create two peculiar effects for an observer on Earth, which can be exploited for a positive identification of a DM signal. Since, in its rotation around the Sun, the Earth's orbital velocity is anti-parallel to the DM wind during summer and parallel during winter, the~observed DM rates inside the detector are expected to display a seasonal modulation of a few percent. A~much more robust signature is provided by the diurnal directional modulation of the DM signal. The~peak flux, in~fact, is expected to come from the direction of solar motion around the center of our galaxy, which happens to point towards the Cygnus constellation. Due to the Earth's rotation around its axis (oriented at 48$^{\circ}$ with respect to the direction of the apparent DM wind), an~observer on Earth would see the average incoming direction of DM changing by $\sim$96$^{\circ}$ 
 every 12 sidereal hours. The~amplitude of the direction modulation depends on the relative angle between the laboratory frame and the Earth axis, with~the maximum at 45$^{\circ}$ inclination and no modulation along directions parallel to the~axis.

The determination of the incoming direction of the WIMP can therefore provide useful information to infer a correlation with an astrophysical source~\cite{Mayet:2016zxu} that no source of background can mimic. Directional measurements can furthermore discriminate between various DM halo models and provide constraints on WIMP properties, unlike conventional non-directional detectors~\cite{Mayet:2016zxu}.

While the last few decades have seen enormous advances in direct DM searches, leading to many orders of magnitude of improvement for masses larger than 10 GeV, the~O(GeV) mass range still remains theoretically well-motivated~\cite{bib:zurek, bib:petraki, bib:relic}. Despite the great effort devoted to lowering the threshold for nuclear recoils to include DM scattering directly from electrons~\cite{bib:essig, bib:essig2, bib:agnes, bib:sensei, bib:supercdms, bib:xenon} or to exploit new signatures such as the Migdal effect~\cite{bib:bernabei,bib:migdal1,bib:migdal2,bib:migdal3, bib:GrillidiCortona} and photon bremsstrahlung~\cite{bib:4}, the~O(GeV) mass range is still largely~unexplored.

Given the kinematics of elastic scattering, a~direct DM detection experiment achieves its best sensitivity for WIMP masses equal to the target mass nuclei. The~maximum fraction $\epsilon$ of the energy that can be transferred to a target of mass $m_T$ by a WIMP of mass $m_{\chi}$ is in fact given by:
\begin{equation}
\epsilon = \frac {4 \rho}{\left( \rho + 1 \right)^2}
\label{eq:eps}
\end{equation}
with $\rho = \frac{m_T}{m_{\chi}}$. Therefore, low-mass target nuclei, such as hydrogen and helium, are the best choices to maximise the sensitivity to O(GeV) WIMP~masses. 

The CYGNO experiment proposes an innovative approach to the direct DM search challenge. A~high-resolution 3D gaseous time projection chamber (TPC) operated at atmospheric pressure is employed with low-mass target nuclei, such as helium and fluorine, to~boost the sensitivity to O(GeV) WIMP masses for both spin- independent (SI) and spin-dependent (SD) couplings. It is also important to notice that low mass nuclei will result in longer track lengths, easing the determination of their direction and thus increasing directional sensitivity. Studies to add a hydrogen-based gas to provide even lighter targets are ongoing.
The topological signature of the recoil event also improves particle identification and hence rejection of natural radioactivity backgrounds, in~particular electron recoils (ER) produced by photon interactions down to low energy thresholds. The~possibility of operation at atmospheric pressure  guarantees a reasonable volume-to-target-mass ratio, while at the same time allowing for a reduction in the engineering requirements of the vessel (hence internal backgrounds). 
The possibility of a high resolution 3D TPC, such as the one foreseen by this project,  will allow CYGNO to explore new physics cases including, among~others, the~elastic scattering of sub-GeV DM~\cite{Baracchini:2020owr} and of solar neutrinos~\cite{Seguinot:1992zu,Arpesella:1996uc}. Moreover, operating at atmospheric pressure would allow a high-mass target to be exposed without the need of very large~volumes.

The results obtained with current prototypes (Section \ref{sec:results}) 
are the basis for the design of a 1 m$^3$ demonstrator that is the subject of this paper. 
According to the performance of this, the~collaboration will propose a larger detector for a competitive experiment.
With this program, CYGNO fits in the context of the wider international CYGNUS effort to~establish a galactic directional recoil observatory that can test the DM hypothesis beyond the neutrino floor and measure the coherent scattering of neutrinos from the Sun and supernovae~\cite{Vahsen:2020pzb}.

\section{The Experimental~Approach}\label{sec:project}

The CYGNO experiment goal is to deploy at INFN Gran Sasso Laboratories (LNGS) a high-resolution TPC with optical readout based on gas electron multipliers (GEMs) working with a helium/fluorine-based gas mixture at atmospheric pressure for the study of rare events with energy releases in the range between hundreds of eV up to tens of ~\keV.

Although challenging, gaseous TPCs constitute a promising approach to directional DM searches providing a set of crucial~features:
\begin{itemize}

\item TPCs usually comprise a sensitive volume, filled with gas or liquid, enclosed between an anode and a cathode generating a suitable electric field in it~\cite{bib:tpc1, bib:tpc2, bib:tpc3}. The~passage of an ionising particle produces free electrons and ions that start to drift towards the above-mentioned electrodes. These are usually segmented and read out to provide granular information about the charge collection point on the plane. The~third coordinate can be evaluated from the drift time measurement.
Therefore, TPCs are inherently 3D detectors capable of acquiring large sensitive volumes with~a lower amount of readout channels with respect to other high-precision 3D tracking~systems;

\item Gaseous detectors can feature very low-energy detection thresholds. A~single electron cluster can be produced with energy releases of the order of few tens of eV and, in~gases, this has a very good chance of reaching the multiplication region to produce a detectable~signal;

\item A measurement of the  total ionisation indicates the energy released by the recoil, and (depending on the readout plane granularity) the profile of the energy deposit along the track can be measured with high precision, providing excellent background~discrimination;

\item Depending on the energy and mass of the recoiling particle and on the gas density, the~track itself indicates the axis of the recoil, and the charge profile along it encodes the track orientation ({\it head-tail}), providing an additional powerful observable for DM~searches;

\item  A large choice of gasses can be employed in TPCs, including light nuclei with an odd number of nucleons (such as fluorine),  which are also sensitive to both SI and SD interactions in the O(GeV) mass~region;

\item A room-temperature and atmospheric-pressure detector results in operational and economical advantages, with~no need for cooling or vacuum sealing. These choices allow for a simpler technology and experiment realization and more straightforward scaling when compared to cryogenic solutions currently dominating the DM direct search~scene;

\item The use of a gaseous target reduces the interaction probability with respect to denser material (liquid or solid). Nevertheless, TPCs up to 100 m$^{3}$ of active volume have already been successfully operated~\cite{bib:alice, bib:tpc4}, 
showing the feasibility of very large detectors with large active masses.
\end{itemize}

\subsection{The Optical~Readout}
\label{sect:opro}

Gas luminescence is a well-studied and established mechanism: charged particles traveling in the gas can ionize atoms and molecules but can also excite them. During~the de-excitation processes, photons are emitted. The~amount and spectrum of light produced strongly depends on the gas, on~its density and on the possible presence and strength of an electric field. In~most common gas mixtures, the~number of emitted photons per avalanche electron can vary between $10^{-2}$ and $10^{-1}$~\cite{bib:Fraga, bib:Margato2, bib:lumi, bib:monte}.

The idea of detecting the light produced during the multiplication processes, proposed many years ago~\cite{bib:charpak}, has received renewed attention in recent years. The~optical readout approach, in~fact, offers several~advantages:
\begin{itemize}
\item Highly performing optical sensors are being developed for commercial applications and can be easily procured;
\item Light sensors can be installed outside the sensitive volume, reducing the interference with high-voltage operation and gas contamination;
\item The use of suitable lenses allows the possibility of imaging large O(1) m$^2$ areas with a single sensor while maintaining an O(100) $\mu$m effective pixels transverse size.
\end{itemize}

In recent years, an~increasing number of tracking detectors have started employing 
micro-pattern gaseous detectors (MPGDs). Their major advantages are their very high achievable granularity and rate capability, together with mechanical robustness. The~production technology for MPGDs nowadays guarantees very high-quality devices, providing stable and uniform operation. In~particular, GEMs~\cite{bib:gem} have already been used to equip very large areas with high spacial and time resolution~\cite{bib:alice}, and~have more recently been employed coupled to pixelised light sensors, showing very good performances~\cite{bib:ref1, bib:jinst_orange1, bib:loomba, bib:Fraga}.

Charge coupled devices (CCD) have been widely used in the past as high granularity light sensors for optical TPC approaches \citep{bib:ccd1, bib:loomba, bib:ccd2}. CCDs' main limitation for the study of rare events in the 1-100~\keV energy range is represented by the high level of readout noise, up~to 5 to 10 electrons RMS per pixel. More recently, cameras based on active pixel sensor (APS) technology developed on complementary metal-oxide semiconductors (CMOS) have been developed that can reach tens of millions of pixels with sub-electron readout noise and single photon sensitivity (usually referred as {\it scientific CMOS} (sCMOS). 

The CYGNO collaboration proposed the introduction of sCMOS-based optical devices for GEM readout in 2015~\cite{bib:nim_orange1}. The~high sensitivity of this technique resulted in a very good performance in particle detection not only at the energies of interest for DM searches (as is illustrated in Section~\ref{sec:results}), but~also for minimum ionising particles from~both cosmic rays and high energy electrons~\cite{ bib:jinst_orange1, bib:jinst_orange2, bib:ieee_orange, bib:elba, bib:lemon_btf}.

Because the current frame rate available for CCD or sCMOS is still low compared to the temporal extent of typical TPC signals, such devices can provide only 2D projection of the recoil track. In~order to achieve 3D track reconstruction, the~CYGNO experiment aims at complementing the sCMOS image information with the signal of a fast light sensor (PMT or SiPM) that can provide the track profile along the drift~direction.

\subsection{The Gas~Mixture}\label{sec:gas}

The relative photon yield, defined as the ratio between the number of produced photons and the total number of secondary electrons produced in the avalanche process, and~in general the overall detector performances, are significantly dependent on the gas characteristics: ionization statistics, transport properties (drift velocity and diffusion), electron multiplication and light production. In~the context of optical TPCs for DM searches, CF$_4$ is a particularly interesting gas because of its well-known property as an efficient scintillator. Furthermore, the large fluorine content provides sensitivity to spin-dependent WIMP-proton interactions. It was demonstrated in previous studies~\cite{bib:Fraga} that CF$_4$-based mixtures have electro-luminescence emission spectra with a large peak around 600~nm, where Si-based sensors (CCD or sCMOS) offer their highest quantum~efficiency.

\begin{figure}[H] 
 
\includegraphics[width=0.35\textwidth]{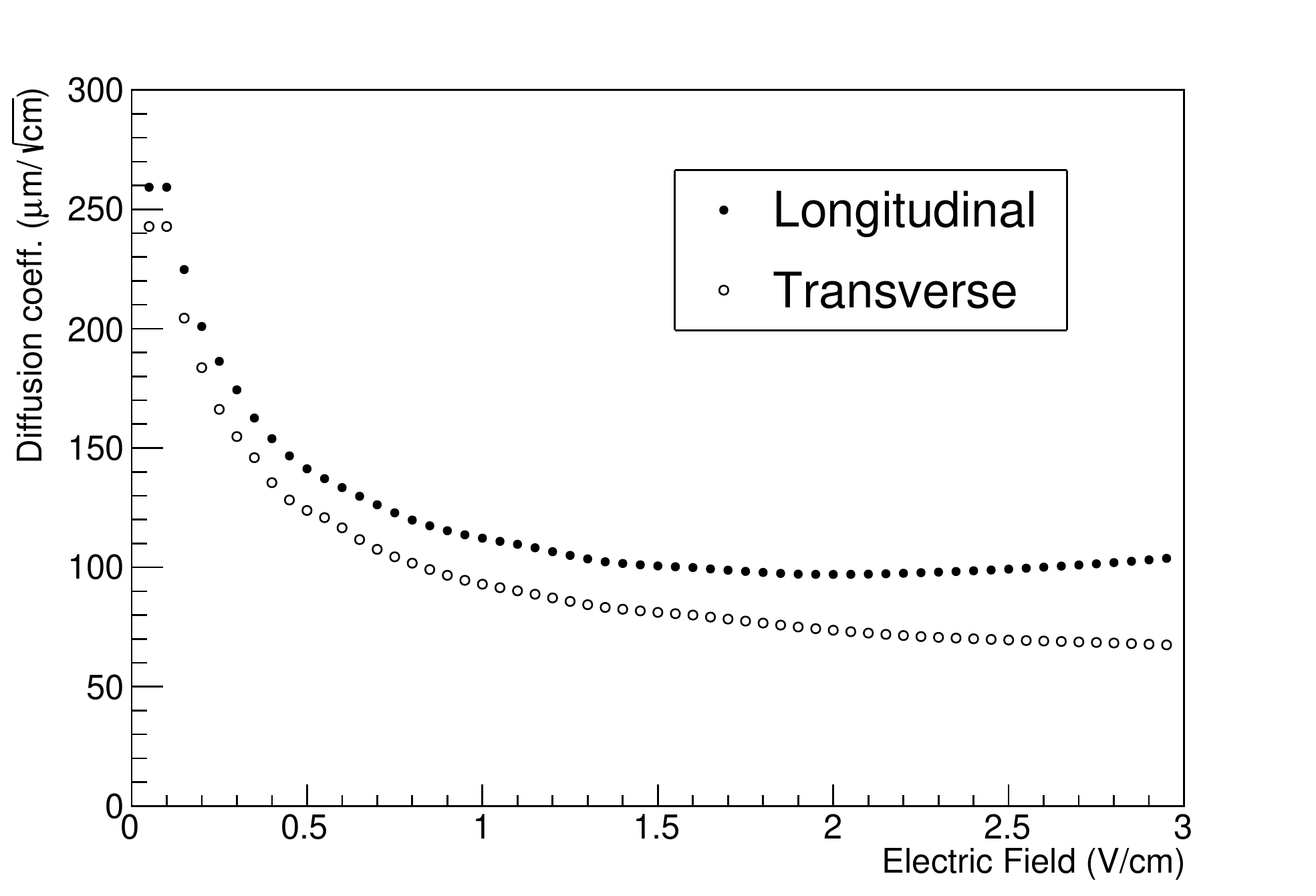}
\includegraphics[width=0.35\textwidth]{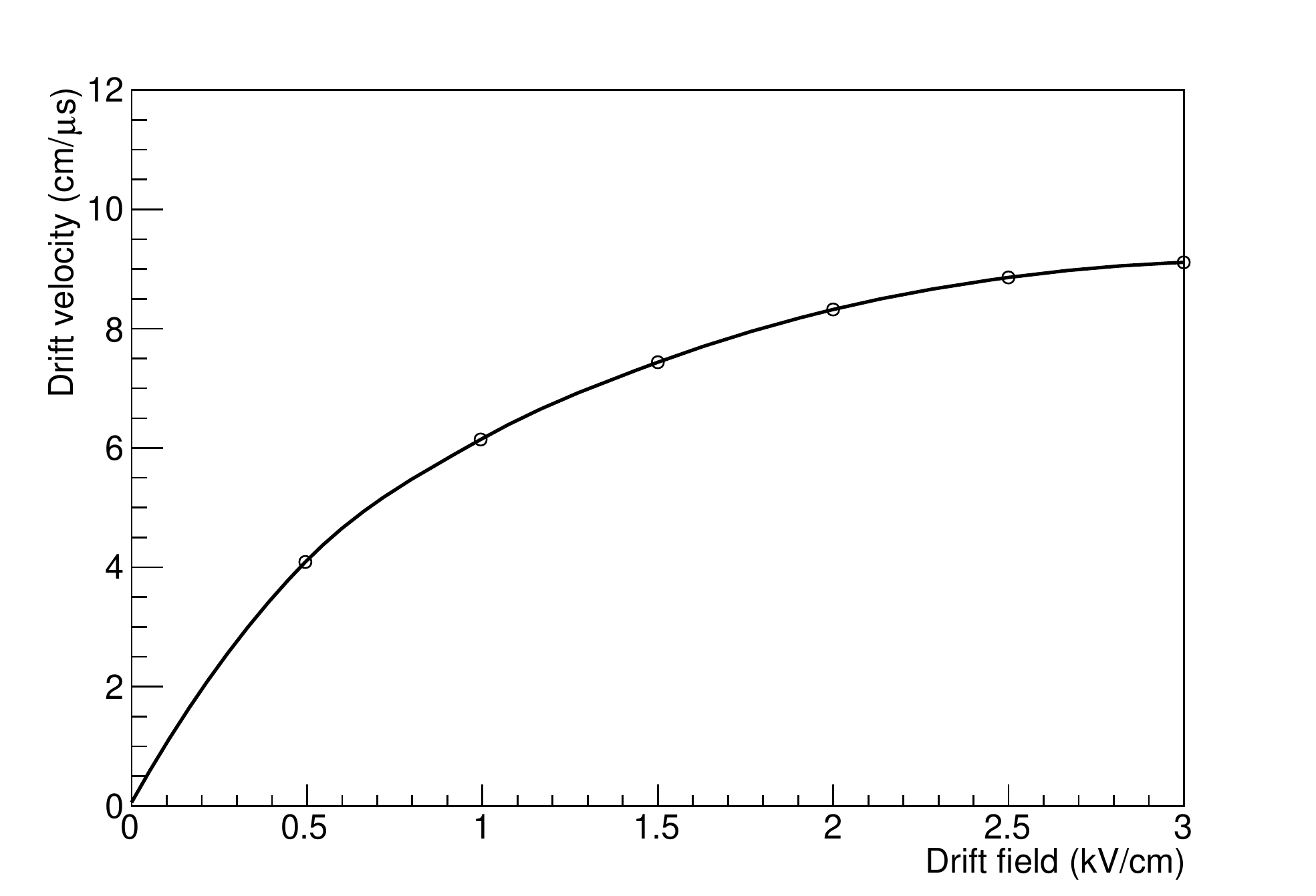}
\caption{Transverse and longitudinal diffusion coefficients for He/CF$_{4}$ 60/40 (\textbf{left}) and electron drift velocity as a function of the drift field (\textbf{right}).}
\label{fig:diff_vdrift}
\end{figure}

For these reasons, He/CF$_{4}$ mixtures in different proportions were extensively studied within the CYGNO project. While the light yield increases for larger amounts of helium, the~detector electrical stability improves for higher CF$_{4}$ contents. As explained in~\cite{bib:fe55New, bib:roby}, the~best compromise  was found for a mixture with 60\% helium and 40\% CF$_{4}$. The~behaviors of the diffusion coefficients and drift velocity for different electric fields were calculated with Garfield~\cite{bib:garfield1,bib:garfield2} and are shown in Figure~\ref{fig:diff_vdrift}.

As can be seen from Figure~\ref{fig:diff_vdrift}, a~remarkable additional advantage of the use of CF$_4$ is the small electron diffusion, which can provide a reduced deterioration of the track's original~shape.

\begin{figure}[H] 
\includegraphics[width=0.7\textwidth]{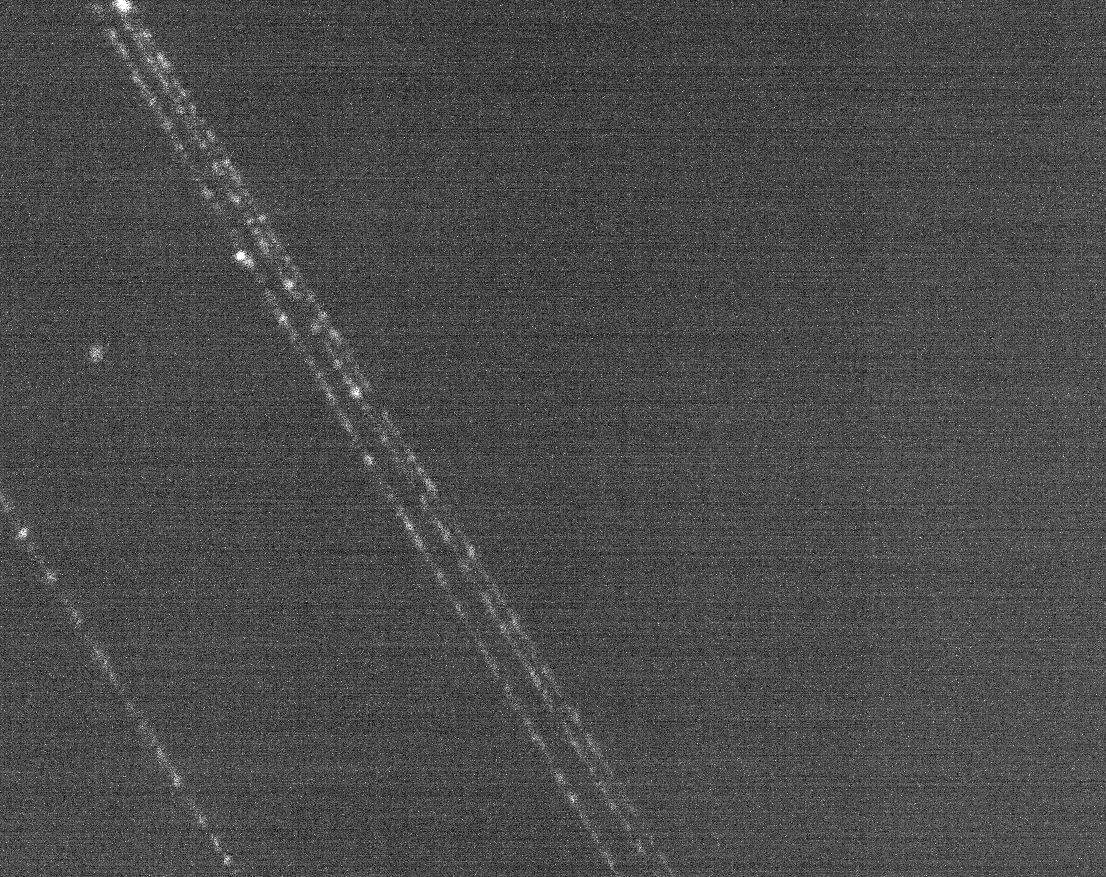}
\caption{Detail of an image collected with the sCMOS sensor of several tracks in cosmic~rays.}
\label{fig:mu_str}
\end{figure}

For this mixture, an~average energy loss per single ionization of 42~eV was estimated.
For a minimum ionising particle, 3 ionizations per track millimetre are expected with about 2 electrons per cluster, resulting in an energy loss of about 250~\eV/mm. An~example of an image of a few cosmic rays is shown in Figure~\ref{fig:mu_str}: ionization clusters are well-visible along the~tracks.

\begin{figure}[H] 
    \includegraphics[width=0.8\linewidth]{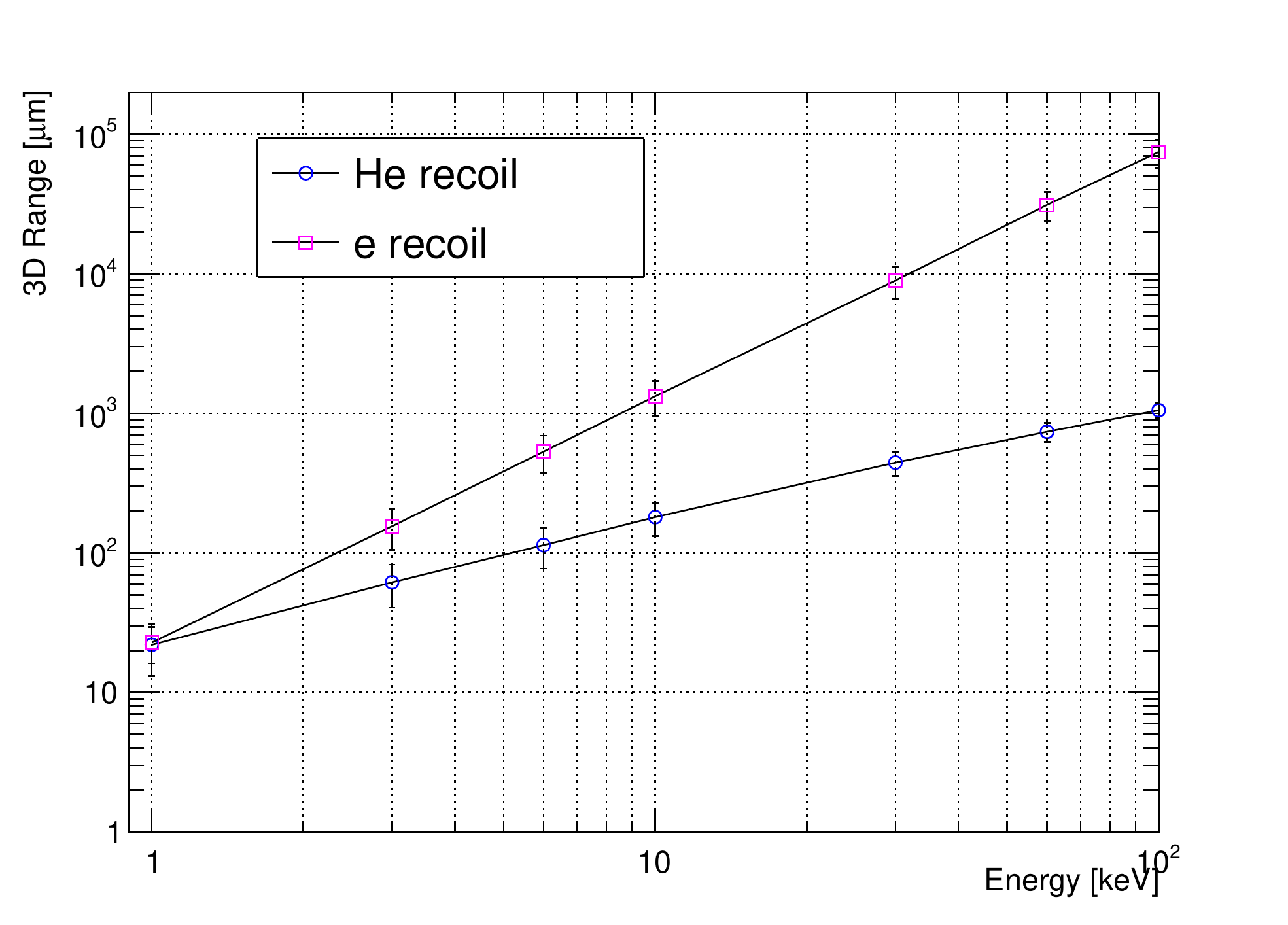}
    \caption{Average 3D distance between the production and absorption point for electron- and He-nucleus recoils as a function of their kinetic energy in a He/CF$_4$ (60/40) gas~mixture.}
      \label{fig:range}
    
\end{figure}

The effective ranges of electron and He-nuclei recoils were simulated,
respectively, with the \GEANT~\cite{bib:geant} and {\it Stopping and Range of Ions in Matter}
 (\SRIM)~software (Visit the \url{http://www.srim.org/} site for more information, accessed on 18 November 2021).
  The~average 3D ranges (i.e., the distance between the production and absorption point) as a function of the particle kinetic energy are shown in Figure~\ref{fig:range}:
 \begin{itemize}
     \item He-nuclei recoils have a sub-millimetre range up to energies
       of 100\keV and are thus expected to produce bright spots with
       sizes mainly dominated by diffusion effects;
     \item Low-energy (less than 10\keV) electron recoils are, in
       general, larger then He-nuclei recoils with the same energy and are
       expected to produce less intense spot-like signals. For~a kinetic
       energy of 10\keV, the~electron range becomes longer than
       1\unit{mm}, and for a few tens of \keV, tracks of a few centimetres
       are expected.
 \end{itemize}
\vspace{-16pt}

Measurements with other mixtures are also being carried out, including, in~particular, the~possible use of small amount of hydrocarbons (e.g., C$_4$H$_{10}$ or CH$_4$) to add protons as low mass targets.  Additionally, alternative gases with reduced global warming power gases (e.g., HFO) are being~tested.

\section{Experimental Results with \lemon~Prototype}\label{sec:results}
\begin{figure}[H]
 
\includegraphics[width=0.7\textwidth]{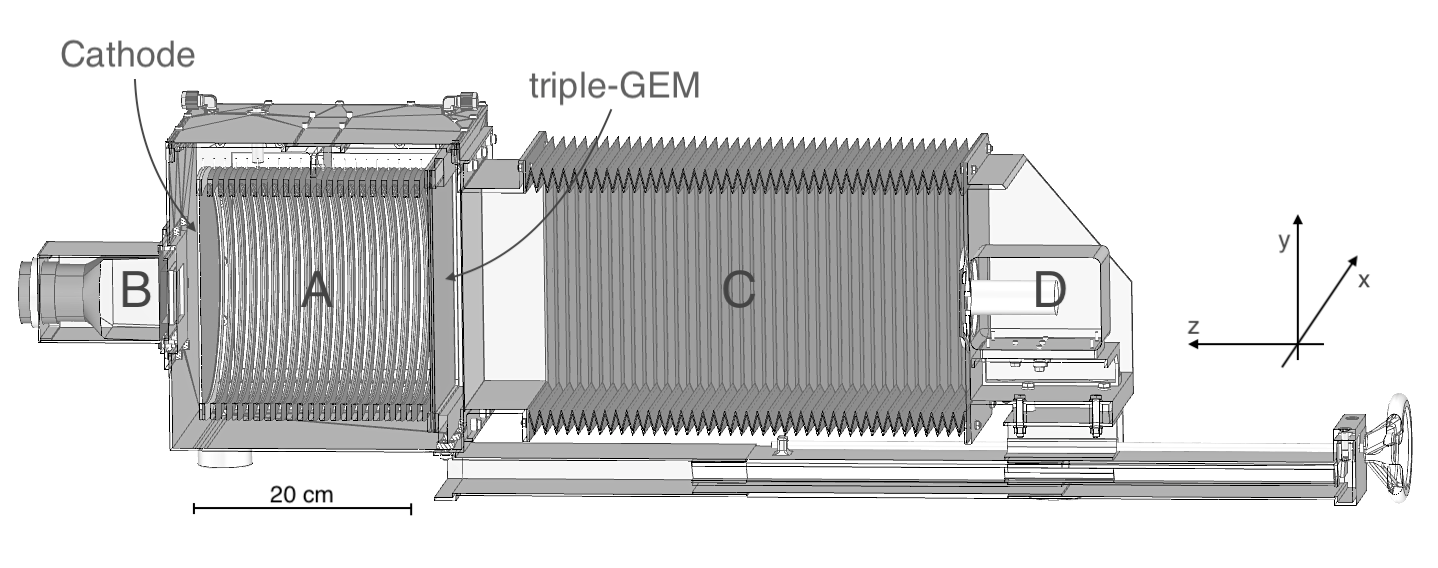}
\caption{The \lemon\ {prototype}~\cite{bib:Antochi_2021}. The~elliptical sensitive volume (\textbf{A}), the~fast photo-multiplier (\textbf{B}), the~optical bellow (\textbf{C}) and the sCMOS-based camera (\textbf{D}) are~indicated.}  
\label{fig:lemon}
\end{figure}

The experimental results obtained with the prototype named the Long Elliptical MOdule (\lemon) represents the most comprehensive example currently available of the performance achievable with the CYGNO approach.
The ~\lemon\ detector,  shown in the schematic of Figure~\ref{fig:lemon}, is composed~of:
\begin{itemize}
    \item A gas sensitive volume of 7 litres contained in a 20~cm long cylindrical field cage (FC) with an elliptical base with 24~cm and 20~cm axes [A];
    \item A 24 $\times$ 20~cm$^2$ stack of 3 GEMs as the amplification stage  facing the sCMOS camera [D], optically coupled through a 50~cm long (see Equation~(\ref{eq:demagnification})) black bellow [C] to protect the optics from external light, with the bottom electrode of the last GEM used as the anode; 
   \item A mesh-based semitransparent cathode closing the volume on the opposite side, behind~which a PMT [B] is placed.
\end{itemize}
A more detailed description of this prototype can be found in Ref.~\cite{bib:fe55,bib:lemon_btf}.

\lemon\ standard operating conditions were based on the following~sets: 
\begin{itemize}
    \item An He/CF$_4$ (60/40) gas mixture flux of 200 cc/min;
    \item An electric drift field within the sensitive volume \Ed~=~0.5~kV/cm;
    \item An electric transfer field in the 2~mm gaps between the GEMs \Et~= 2.5~kV/cm;
    \item A voltage difference across the two sides of each GEM \Vg~=~460~V;
\end{itemize}

According to results presented in~\cite{bib:roby}, in~this configuration, an electron multiplication of about 1.5$\times 10^6$ is~expected.

As anticipated in Section~\ref{sect:opro}, high-quality cameras are a crucial ingredient for the experiment results. As~a result, an~ORCA Flash 4.0 camera ({for more details visit \url{www.hamamatsu.com}}) was selected to equip \lemon. This device is based on a 1.33~$\times$~1.33~cm$^2$  sCMOS sensor, subdivided in 2048~$\times$~2048 pixels with an active area of 6.5~$\times$~6.5~$\mu$m$^2$ each, with~a quantum efficiency of 70\% at 600~nm and a readout noise of 1.4~electrons RMS. The~response and noise level of this sensor were tested with a calibrated light source~\cite{bib:jinst_orange1}. A~response of 0.9 counts/photon was measured together with a RMS fluctuation of the pedestal of 1.3 photons/pixel. 
 
In order to image the large GEM surface, the~camera is equipped with a Schneider lens with a 25.6~mm focal length $f$ and a 0.95 aperture $a$. Since at a distance $d$ the lens provides a de-magnification of
\begin{equation}
\label{eq:demagnification}
 \delta = \frac{f}{d-f} 
\end{equation}
the camera optical system is placed at $d$ = 52.6~cm distance from the GEMs in~order to image a 26 $\times$ 26~cm$^2$ area. The~solid angle covered by the sensor, which in turn determines the geometrical acceptance of photons, is given~by
$$
\label{eq:omega}
\Omega = \frac{1}{\left(4(1/\delta+1)\times a \right)^2}
$$
resulting in $1.6 \times 10^{-4}$ for the \lemon\ layout.

In order to complement the 2D track projection recorded by the sCMOS with the track trajectory along the drift direction, the~arrival time profile of the primary electrons could be extracted from the signal induced on the third GEM bottom electrode. Nonetheless, this is expected (and explicitly shown in~\cite{bib:jinst_orange2}) to suffer from considerable
noise (typically due to jitter on the high voltage
supply line) that could prevent signal detection at the low energies at play.
To overcome this limitation, a light track time profile was concurrently readout by a Photonics XP3392 Photo Multiplier Tube (PMT), each with a 5~ns rise-time, a~maximum QE of 12\% for 420~nm and a 76~mm square-window, providing sensitivity to a single photon and significantly reduced noise with respect to the GEM electric~signal. 


The performances
of \lemon\ has been tested in recent years at the INFN Laboratori Nazionali di Frascati (LNF) overground laboratory by means of radioactive sources (\fe, \ambe), high-energy (450 MeV) electrons from a beam at the Beam Test Facility (BTF,~\cite{bib:btf1,bib:btf2}) and cosmic rays, and~are summarised in the~following.

\subsection{Operation~Stability}\label{sec:stability}

The performance and long-term stability of \lemon\ was
tested for a month long run, during~which the detector was exposed to environmental radioactivity, cosmic rays and a \fe~source~\cite{bib:fe55New}. During~the whole period, all currents drawn by the high-voltage channels supplying the electrodes of the GEM stack were monitored and recorded to identify sudden and large increases that could indicate discharges or other electrostatic issues.
During the test, two different kinds of electrostatic instabilities were observed:

\begin{itemize}
    \item Hot-spots appearing on the GEM surface. While in some cases these would fade out with time, sometimes they started to slowly grow up to tens of nA (on a time scale of minutes). These are very likely due to self-sustaining micro-discharges happening in one or a few GEM~holes;
    
    \item High charge density due to very high ionizing particles or charge accumulation on electrode imperfections can suddenly discharge across GEM holes. In~these events, a~sudden increase in the drawn current is recorded with a voltage restoring on the electrodes through $10~\Omega$ 
     protection resistors on a time basis of a few seconds.  Even if these events are less frequent than hot spots, they can be dangerous for the GEM structure and the energy released in the discharge can, in~principle, damage it. 
\end{itemize}

An automatic recovery procedure was implemented, triggered by the raising of the GEM currents, which was able to recover both hot-spots and discharges by lowering and gradually restoring the GEM voltage operating conditions in a few~minutes.

An average of 16 such instabilities per day were observed and the total dead time introduced by the recovering procedures was less than 4$\%$. A~detailed analysis of the time interval between two consecutive phenomena did not show any correlation between two subsequent events, nor any increase of their rates. This study demonstrated that the detector operation looked very safe and stable, and the obtained performance is considered to be satisfactory. Different gas proportions were tested, and a lower amount of CF$_4$ resulted in a less stable electrostatic~configuration. 

\subsection{Light Yield and Energy~Resolution}\label{sec:yield}

The light production was evaluated by analysing the response of sCMOS and PMT to interactions in a gas of 5.9~keV X-rays produced by a \fe~source. The~sCMOS images were acquired in a free-running mode (i.e., without using any trigger signal) with an exposure of 100 ms. The~sCMOS pixels' pedestal noise was extracted from an average of 100 images acquired in the absence of any light signal and subtracted from each image before the analysis. An~elementary clustering algorithm based on the nearest neighbor-cluster (NNC) is applied to 4 $\times$ 4 rebinned images to select \fe\-induced energy~deposits.

\begin{figure}[H]
 
\includegraphics[width=0.35\textwidth]{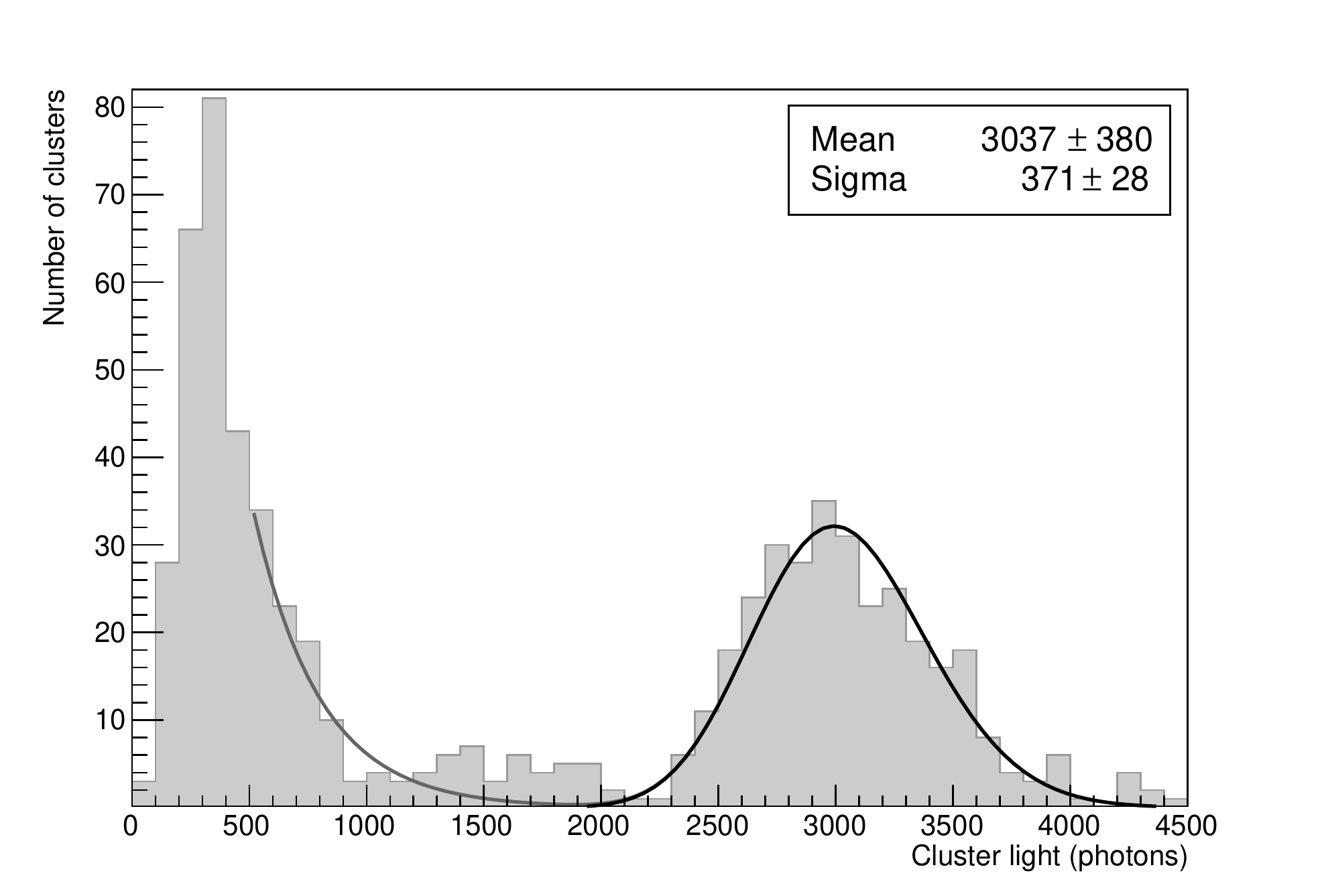}
\includegraphics[width=0.35\textwidth]{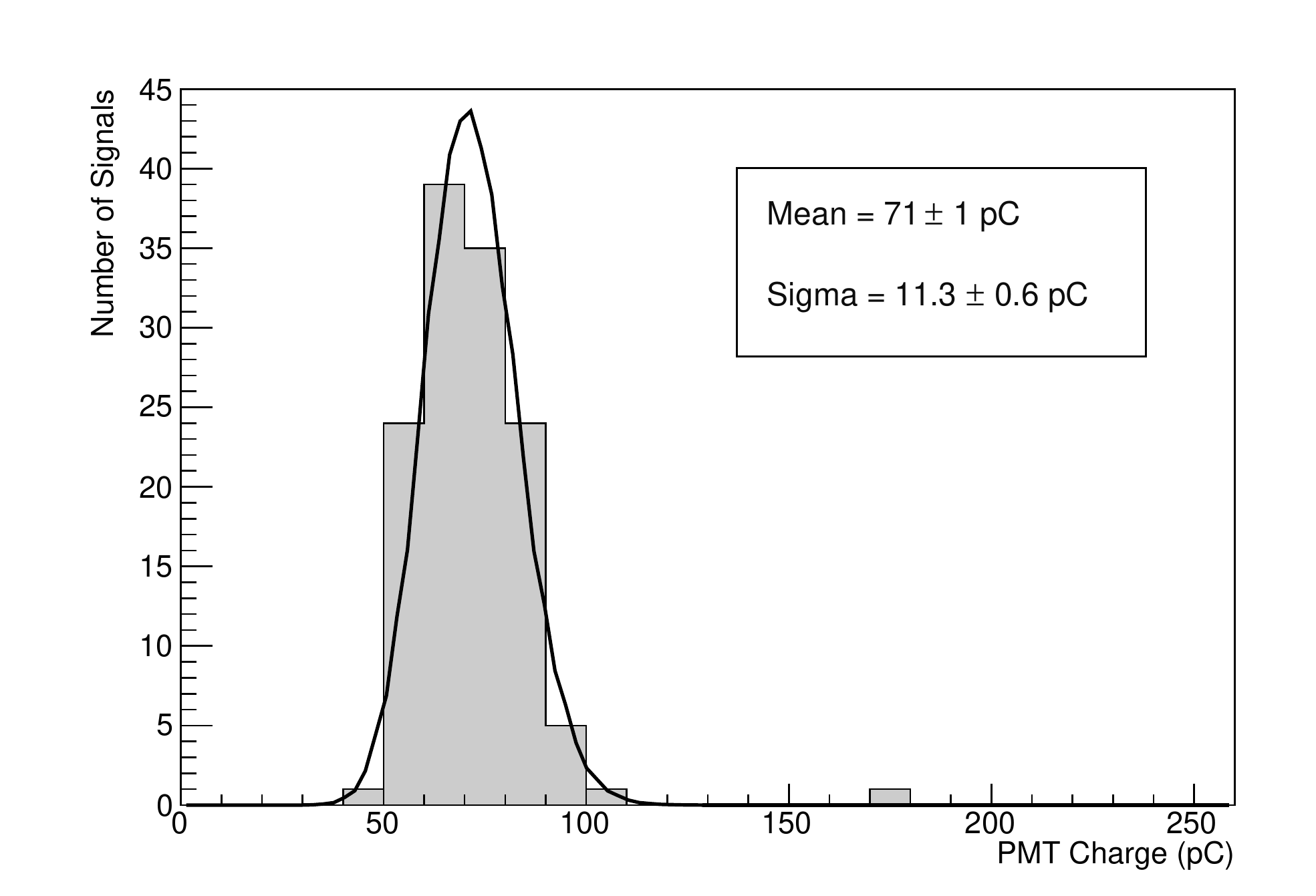}
\caption{Distribution of the light content of the \fe~events reconstructed from the sCMOS images  (\textbf{left}), and distribution of the charge measured by the PMT signals (\textbf{right}).} 
\label{fig:light}
\end{figure}

Figure~\ref{fig:light} shows, on the left, the light spectrum of the \fe\ events reconstructed from the sCMOS images, and on the right, the integral of the charge signal measured by the~PMT.

The average light yields were evaluated from a Polya fit~\cite{bib:rolandiblum} to the two distributions, resulting in an average of 514~$\pm$~63 photons per \keV detected by the sCMOS camera (in agreement with results obtained with lower \Vg\ and \Et\ \cite{bib:fe55}), with~an RMS energy resolution of 12\% and an average of (12.0~$\pm$~0.2) pC per \keV together with an RMS energy resolution of 16\% by the PMT charge~signal.

The energy resolutions are mainly due to the Poisson's fluctuations of the numbers of primary electrons (8\%) and of the gain of the first GEM.
The latter term can be simply evaluated to be about 10\% by supposing an exponential distribution for it~\cite{bib:thesis}, with~an average value of~100.

\subsection{Detection~Efficiency}
\begin{figure}[H]
 
\includegraphics[width=0.35\textwidth]{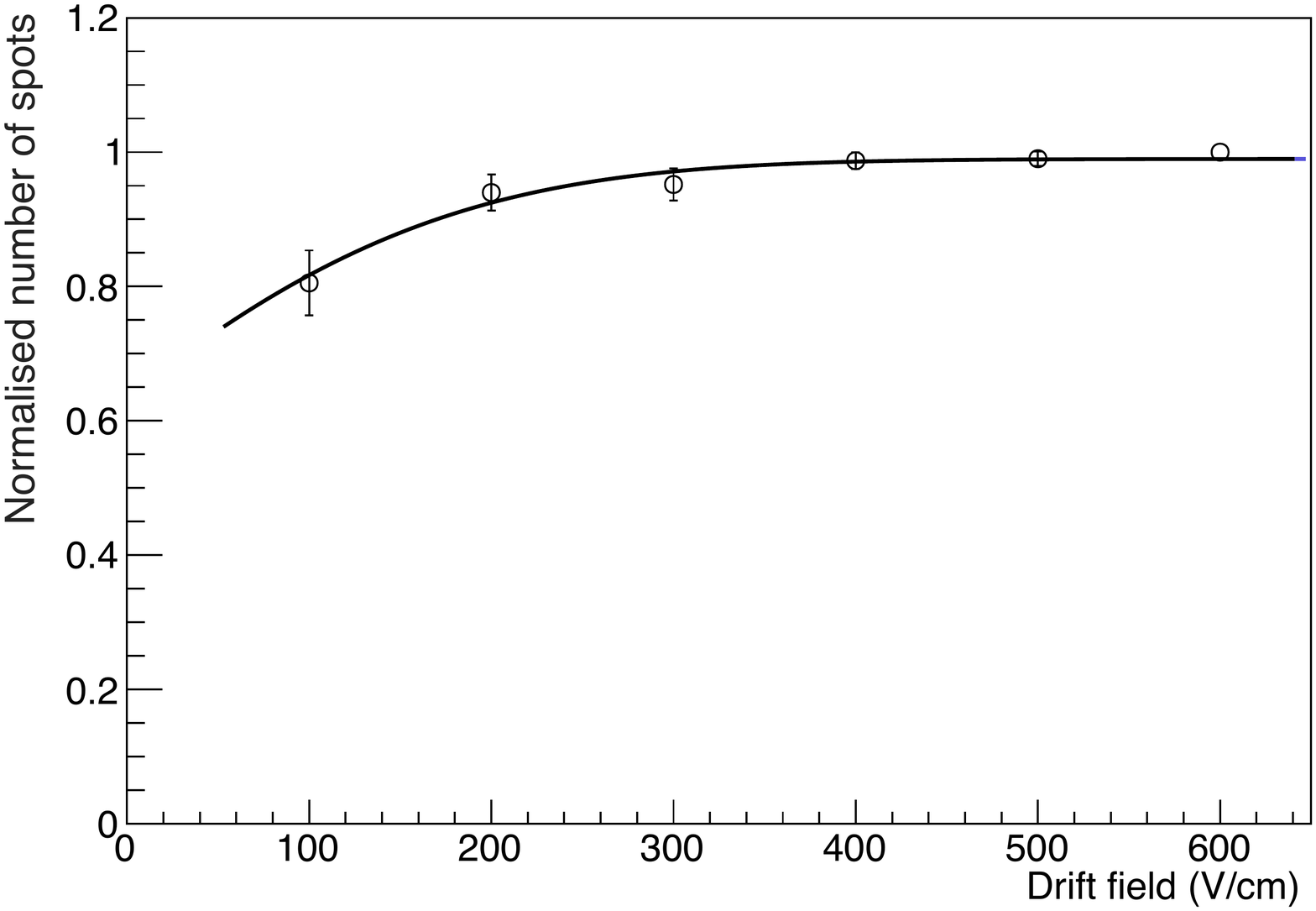}
\includegraphics[width=0.35\textwidth]{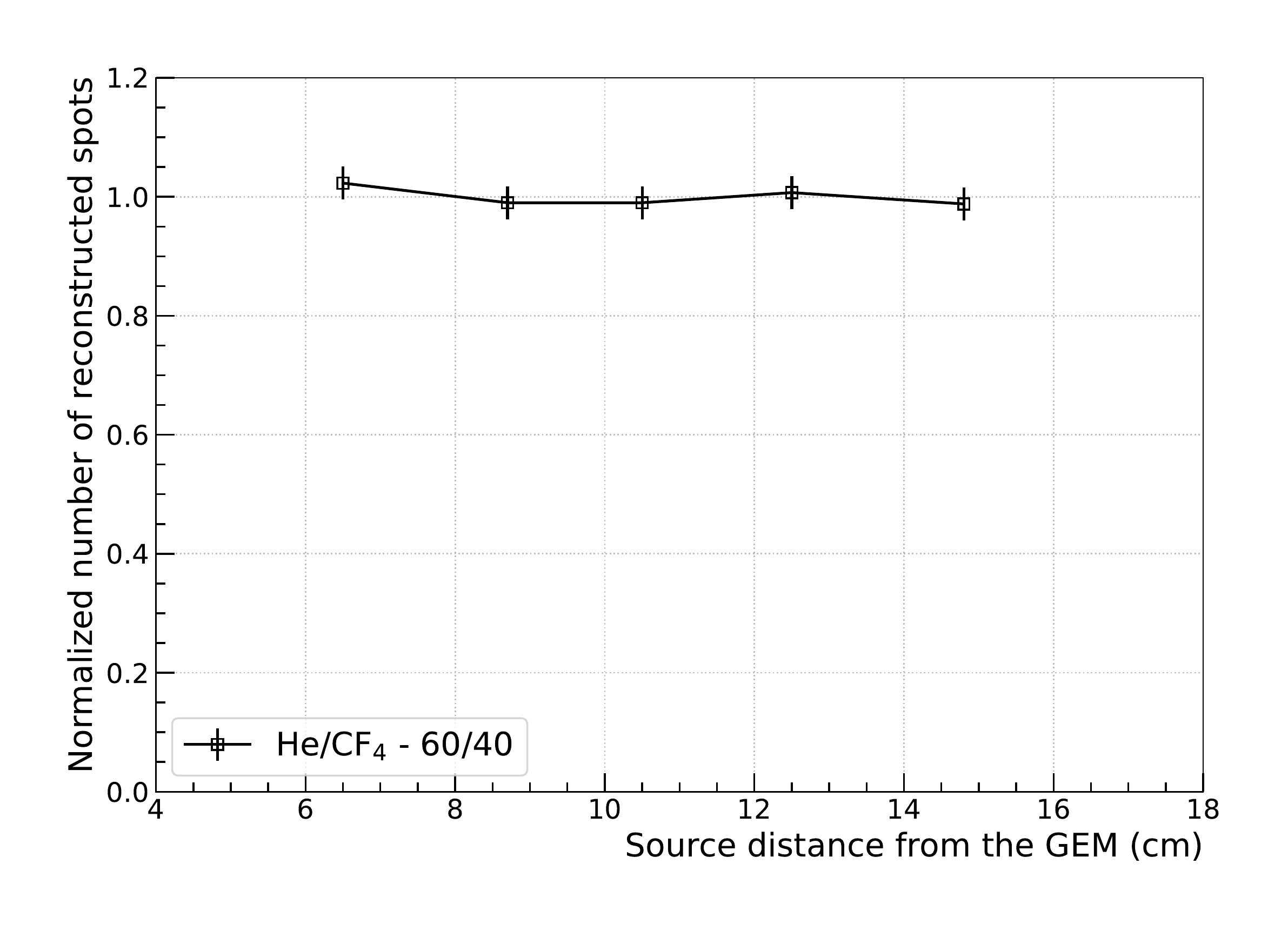}
\caption{Behavior of the normalized number of \fe~spots as a function of the drift electric field (\textbf{left}) and event depth in the sensitive volume (\textbf{right}).} 
\label{fig:deteff}
\end{figure}

The detection efficiency along the whole 7~litre sensitive volume was studied
acquiring sCMOS images by changing the electric field
strength within the field cage (drift field) and the position of a collimated \fe\ source in~order to vary the X-ray interaction distance to the amplification region.
 Figure~\ref{fig:deteff} shows, on the left, the number of reconstructed \fe\ spots in the sCMOS images with the algorithm illustrated in Section~\ref{sec:yield} as a function of drift field (\Ed), normalized to the value obtained for \Ed~=~600~V/cm. For~\Ed\ larger than 300~V/cm, a plateau is found, suggesting  a full detection efficiency. The~right panel of Figure~\ref{fig:deteff} shows the dependence of the number of reconstructed spots normalised to its average value on the source distance from the GEM amplification plane, as~measured with an \Ed\ of 600~V/cm. A~constant behavior is found in all tested positions, indicating a stable detection efficiency that is not dependent on the interaction point distance from the~GEM.

\unskip

\subsection{Track Absolute Distance along the Drift~Direction}\label{sec:track}
The possibility to determine the absolute $z$ position of the energy deposit exploiting the electron drift was studied with 450 MeV electrons from the LNF-BTF facility~\cite{bib:lemon_btf}. The~ transverse diffusion in the drift gap can, in fact, be used to extract the drift length and thus infer the absolute {\it z} distance at which the track occurred. 
Seven millimetre-long track segments were used to evaluate the detector performance for small energy~releases. 

As described in~\cite{bib:Antochi_2021}, the~light profile transverse to the track direction possesses a Gaussian shape with the total light $L$ being proportional to $\sigma \times$ A (where $\sigma$ is the RMS and A is the amplitude of the Gaussian).
Because of the attachment effect in gas~\cite{bib:att}, the~probability for an electron to reach the GEM stack decreases exponentially with a mean free path $\lambda$ (and thus $L = L_0 e^{-z/\lambda}$). Since $\sigma$ is expected to increase as $\sqrt{z}$ because of diffusion in gas, the~ratio $\eta$, defined as $\sigma$/A, is expected, at~a first order approximation, to~grow quadratically with the drift distance~\cite{bib:lemon_btf}.

Similarly, longitudinal electron diffusion modifies the electron time of arrival on the GEM and thus the time structure of the signal recorded by the PMT. Additionally, in this case, the~ratio $\eta_{PMT} = \sigma_{PMT}/A_{PMT}$ is expected to increase with {\it z}.

\begin{figure}[H]
 
\includegraphics[width=.35\textwidth]{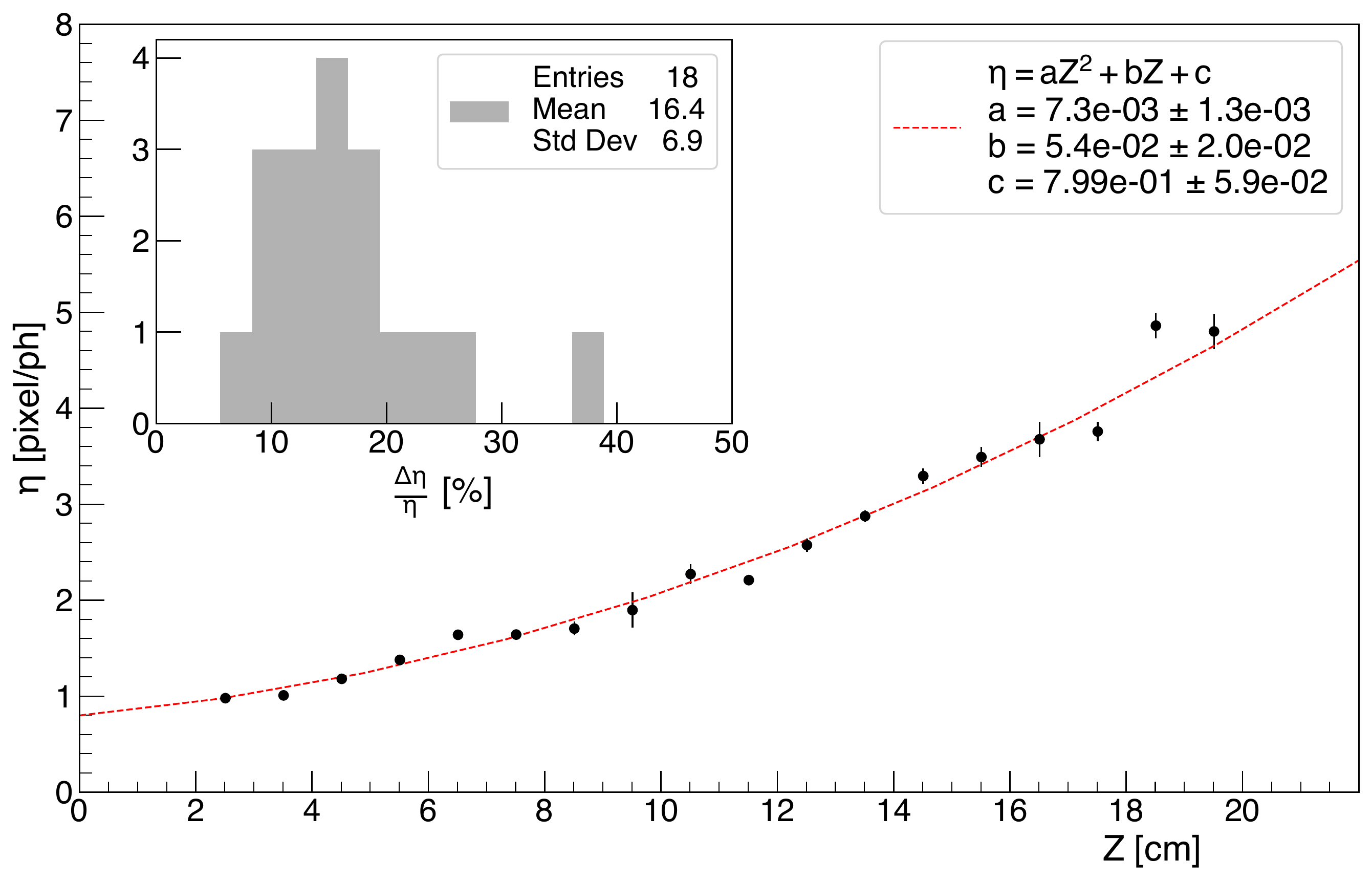}
\includegraphics[width=.39\textwidth]{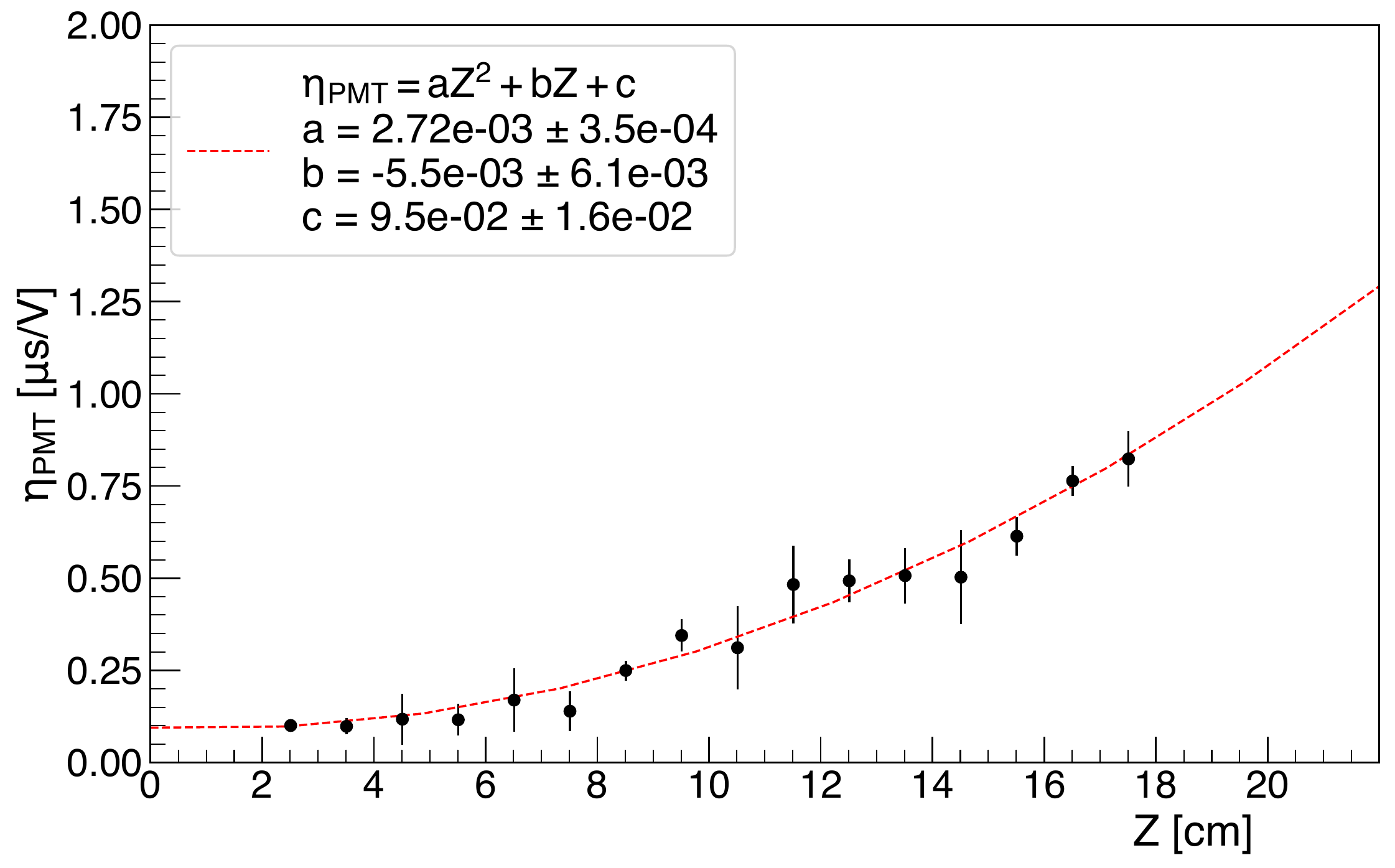}
\caption{Dependence of $\eta$ on the left and $\eta_{PMT}$ on the right as a function of the track distance from the GEM (see text for details).}
\label{fig:eta}
\end{figure}

Figure~\ref{fig:eta} shows the dependence of $\eta$ and $\eta_{PMT}$ as a function of {\it z} with a superimposed quadratic fit. 
The inset shows the distribution of the ratio between the RMS and the average values of the spectra of $\eta$ obtained at the various $z$. 
These observables can be, therefore, used to evaluate the absolute $z$ with about 15$\%$ uncertainty over a 20 cm length~\cite{bib:lemon_btf}.
These features will allow the fiducial signal volume to be selected, therefore rejecting background signals coming from the radioactivity of TPC materials, like the cathode or~GEMs.

\subsection{Detection and Identification of Nuclear and Electron~Recoils}
\label{sect:rej}

Thanks to the detailed information provided by the high granularity optical sensor, track properties like the shape, the~length and width, and~light density, among others, can efficiently be exploited to identify and separate ER which are surely due to background sources from NR which are possible candidates of DM~signals.

To quantify these features within the CYGNO experimental approach, a~track reconstruction and identification algorithm was developed for the analysis of the sCMOS images called iDBSCAN~\cite{Baracchini:2020iwg} and based on an adapted version of the well-known Density-Based Spatial Clustering of Applications with Noise (DBSCAN) \cite{dbscan1996}. The~reconstructed clusters were used as seeds for a superclustering algorithm based on Geodesic Active Contours (\gac~\cite{gac,mgac}), which gathers together sub-clusters of the energy deposits belonging to a single track. The~\gac, exploiting the number of photons in each pixel as a third dimension to the phase space of the points considered, separately identifies clusters displaying different intensity (i.e., energy deposition patterns) which are,~therefore, likely belonging to different classes of particles~interactions.

The performance of the algorithm was studied using 5.9 keV energy deposits from $^{55}$Fe and NR  produced by an \ambe~source~\cite{bib:coronello}. The~59 keV photons produced by \ambe~were nearly completely shielded by a lead shield built around the detector. \textls[-11]{Data were taken overground at LNF and were, therefore, highly contaminated by cosmic ray particle~interactions.}

In order to  select a pure sample of nuclear recoil candidates produced by the interaction of the neutrons
originating from the source and to identify various sources of backgrounds, several cluster shape observables were exploited. Among~these, the~\emph{slimness} ($\xi$) was used to mainly distinguish cosmic rays  and the light \emph{density} ($\delta$) to discriminate electron from nuclear recoils. The~slimness is defined as the ratio of the Gaussian width of the track in the transverse direction over the projected path length. The~density is the ratio of the total number of photons detected by all the pixels gathered in the cluster over the total number of~pixels. 

\begin{figure}[H]
   
    \includegraphics[width=0.78\linewidth]{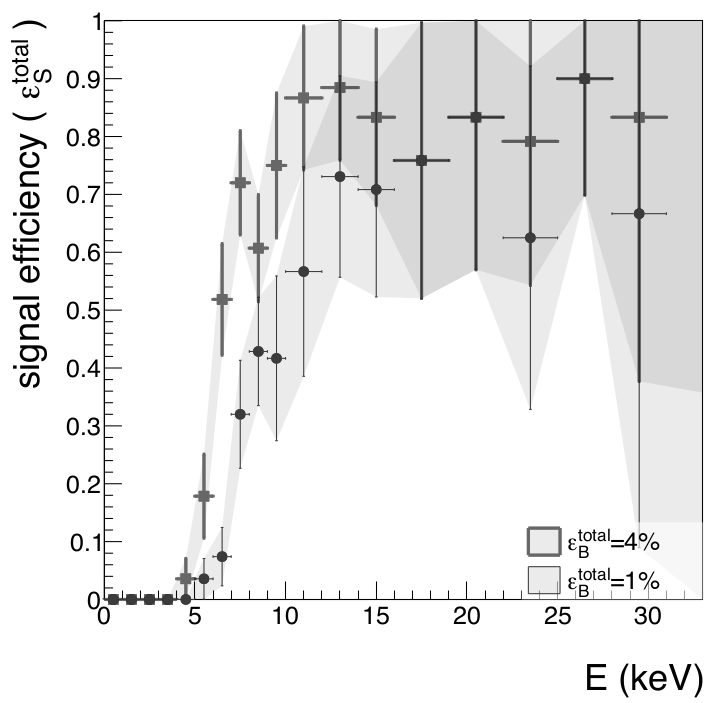}
    \caption{Detection efficiency for nuclear recoils ($\epsilon^{total}_s$) as a function of their detected energy for electron recoils efficiency of 4\% (squares) and 1\% (circles).}
      \label{fig:coronello}
\end{figure}

As shown in Figure~\ref{fig:coronello}, by~simply exploiting a selection on $\delta$, an~ER background rejection in the energy region around 5.9 keV$_{ee}$ of 96.5$\%$ (99.2$\%$) was found together with  50$\%$ (40$\%$) NR efficiency~\cite{bib:coronello}.

While this cut-based
approach is minimalist and~could be improved by more sophisticated analyses combining several topological variables as well as the information from PMT waveforms, it shows
that a rejection factor larger than 10$^2$ for electron recoils at $E=5.9\keV$ can be obtained with a
gas detector at atmospheric pressure while
retaining a high fraction of NR event~signals.

\section{The CYGNO Experiment Roadmap and~Synergies}
The CYGNO project will be developed through a staged approach to~optimise the apparatus and improve its performance while better mitigating any unexpected~cost.

This roadmap, comprises:

\begin{itemize}
    \item PHASE\_0: the installation, in~2022, of~a large prototype (50 litres of sensitive volume) underground at the INFN-Laboratori Nazionali del Gran Sasso (LNGS) to study its performance in a low background environment and validate MC simulation;
    \item PHASE\_1: testing, in~2024--2026, of the~scalability of the experimental approach on a O(1) m$^3$ detector while studying and minimising the radioactivity background due to apparatus material;
    \item PHASE\_2: depending on the results of the previous phases, a larger scale experiment (30--100 m$^3$) will be proposed to explore the 1--10 GeV WIMP mass region with high sensitivity for both SI and SD couplings and the possibility of performing the first measurement of low-energy solar neutrinos.
    In both cases the \emph{directionality capabilities} of the CYGNO approach will allow not only detection of the interactions, but will~also provide useful information for astrophysical studies of incoming particles.
\end{itemize}

\textls[-16]{The roadmap details and synergies with other projects are outlined in the following~sections. }

\subsection{CYGNO PHASE\_0: the LIME~Prototype}\label{sec:phase0}

\begin{figure}[H]
 
 \includegraphics[width=0.7\textwidth]{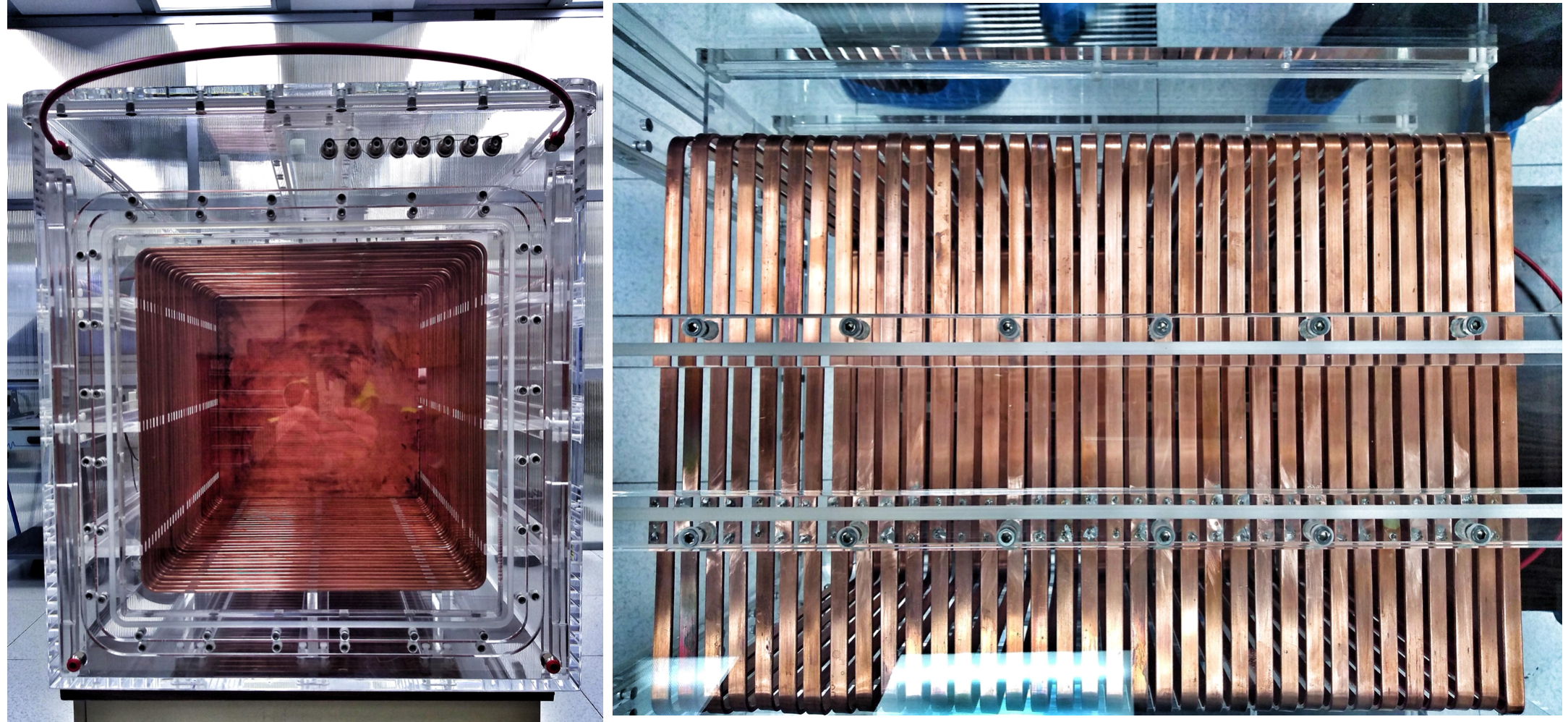}
 \caption{Pictures of the LIME detector. \textbf{Left}: front view of the field cage with the copper cathode visible at the end. \textbf{Right}: field cage copper rings in the gas~vessel.}

 \label{fig:LIME_pic}
 \end{figure}

 The Long Imaging ModulE ({\it \lime}, shown in~Figure~\ref{fig:LIME_pic}), is the larger prototype foreseen to conclude the R\&D phase of the project.
It was conceived to have the same drift length (50~cm) of the final demonstrator (Section~\ref{sec:phase1}) and the same readout scheme based on triple 33 $\times$ 33 cm$^2$ thin GEMs (stretched on a plexiglass frame to reduce radioactivity) imaged by a single sCMOS sensor and 4 small PMT symmetrically placed around the sensor at~a distance of about 15~cm from it and 25~cm apart from the GEM~surface. 

The PHASE\_1 demonstrator will be based on readout modules having the LIME dimensions and layout. For~this reason, its successful assembly and operation will be paramount to substantiate the efforts and confirm the scientific and technological choices towards the 1 m$^3$ detector.

The new Hamamtsu ORCA-Fusion Camera was employed (\url{https://www.hamamatsu.com/eu/en/product/type/C14440-20UP/index.html}, accessed on 18 November 2021) with improved performance with respect to the Orca Flash in terms of reduced noise (0.7 versus 1.4 electrons per pixel), a larger number of pixels (2304 $\times$ 2304 versus 2048 $\times$ 2048) and a larger sensitivity spectrum with a maximum quantum efficiency at 600~nm of 80$\%$ versus 70\%. 
The choice of 4 PMT resides in the possibility of better reconstructing the track position and inclination through the center of gravity of the light signal from the 4 sides. 

The gas volume is enclosed in a 10~mm thick plexiglass box that provides gas tightness. The~field cage is composed by square copper rings, with~a rounded shape to avoid discharges, at~a 16 mm~pitch.

In its underground installation, LIME will be equipped with the same DAQ system and gas system envisaged for the realisation of PHASE\_1, which is currently undergoing~testing.

A response of about 650 ph/keV was measured that has to be compared with 514~ph/keV obtained with \lemon\ (see Section~\ref{sec:yield}) thanks to the larger sensitivity of the Orca Fusion camera. The~low sensor noise (about 1 photon/pixel) will allow operation with an effective energy threshold of hundreds of eV. The~energy resolution on the $^{55}$Fe peak is measured to be 14$\%$ across the whole 50 cm drift length, with~full efficiency in the full 50~litres volume. Furthermore, LIME has already been operated for one entire month, with~its currents continuously monitored and logged, showing comparable stability to \lemon (see Section~\ref{sec:stability}).

The installation at LNGS, completed with the PHASE\_1 auxiliary systems, will~allow:
\begin{itemize}
    \item The detector performance in low radioactivity and a low pile-up configuration to be tested;
    \item The real radioactive background present in the site to be characterized, and then the GEANT4 simulation to be validated.
\end{itemize}

\subsection{CYGNO PHASE\_1: the O(1) m$^3$ Demonstrator}\label{sec:phase1}
Having optimised and assessed all technological aspects with LIME underground tests, the~project will move to PHASE\_1, with~the aim of studying and minimising material radioactivity effects on a real experiment scale, therefore evaluating its~sensitivities.

\begin{figure}[H]
 
 \includegraphics[width=0.7\textwidth]{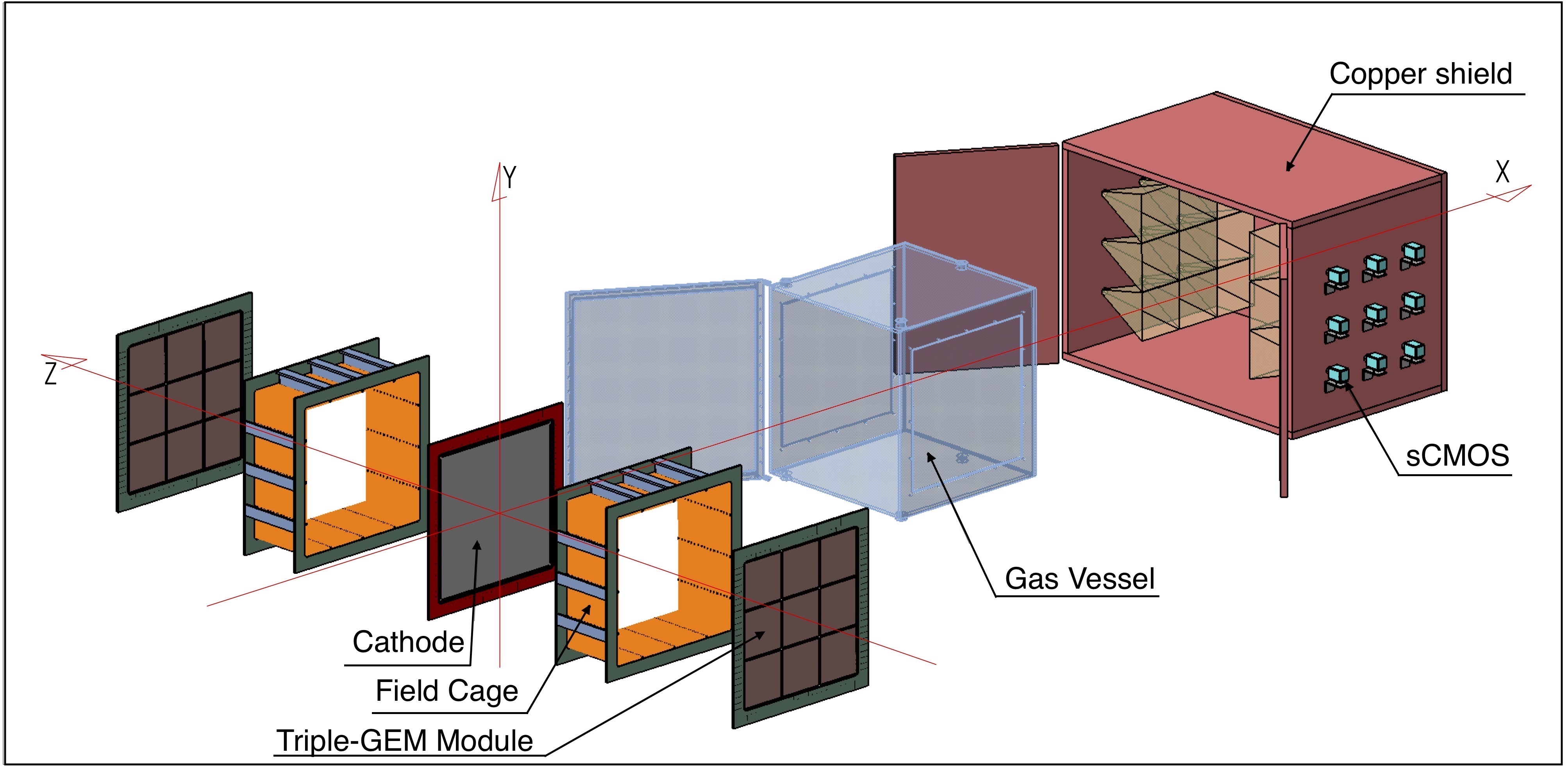}
 \caption{CYGNO PHASE\_1 Detector~Layout.}
 \label{fig:detector}
 \end{figure}

The exact PHASE\_1 detector size will depend on the available underground site, which is still under discussion; regardless, a 1 m$^3$ active volume will be discussed in this paper, schematically shown in Figure~\ref{fig:detector},
with the consideration that the foreseen layout and auxiliary system can be directly and easily adapted to the definitive detector~dimensions.

The active volume of the detector will be contained in a gas volume vessel (GVES) realized with PMMA to lower the material intrinsic radioactivity, reduce the gas contamination, and~ensure the electrical isolation from the cathode and field cage. 
The GVES will contain 2 field cages, 500~mm long, with~2 back-to-back TPCs separated by a central aluminised mylar cathode following the DRIFT example~\cite{bib:drift1, bib:drift2, Battat:2015rna}. This foil is expected to minimise backgrounds induced from recoils by the decay chain of~radon.

Each of the 2 end-caps will have an active area of 1~m$^2$ surface, readout by a matrix of $3 \times 3$ modules of $33 \times 33$ cm$^2$  area. Furthermore, each one will be equipped with a stack of 3 GEMS, 1 sCMOS and 4 PMT, identical to the LIME prototype (see \ref{sec:phase0}).

The GEMs will be assembled adapting the technique discussed in~\cite{Colaleo:2015vsq}. The~mechanical rigidity will be provided by the outer frame that will be anchored to the GVES. 
The assembled GEM stack will be inserted into the GVES through vertical slits, which will allow an easy substitution of a single GEM foil in case of~damage.

The high-voltage system has been conceived with the assumption of independent lines 
for each electrode of the GEM foils in~order to ensure the safe and reliable operation of the detector~modules. 

The DAQ system will be able to collect synchronized data from cameras and photodetectors and to handle the following~specifications: 
\begin{itemize}
    \item Camera exposure from 0.2 to 1 second (1 to 5~Hz frame rate);
    \item 10~MB of data per picture (5~MP, 16 bits/pixel); 
    \item 12-bit digitization of photodetector waveforms at $\sim 250$~MS/s in $\lesssim 1~\mu$s windows.
\end{itemize}
Fast responses provided by the PMT will be exploited to trigger the acquisition of sCMOS sensors. Different possible trigger scheme are under evaluation along with the possibility of running either in trigger or trigger-less~mode.

The acquisition will be distributed through a redundant system of 
machines to ensure the system stability. To~acquire fast photodetectors, digitization boards are considered. In~this scenario, the~bottleneck for the acquisition would be the throughput to the disk, typically limited to O(200~MB/s),
and some preselection of the images by a farm of CPUs would be~needed. 

\subsubsection{PHASE\_1 Shielding Scheme and Material~Budget}\label{sec:back}

A \GEANT~based Monte Carlo simulation of the whole apparatus, shown in Figure~\ref{fig:detector}, has been developed to study detector backgrounds and to optimise the choice of shielding and~materials. 

The effect of the diffused environmental gamma rays and neutron flux was studied. Different configurations of external passive shielding with layers of copper, lead and water were studied with the goal of having less than 10$^4$ photons/year interacting in the target gas between 1 keV and 20 keV. The~choice of this benchmark is backed up by indication from measurements~\cite{Riffard:2016mgw,Phan:2015pda} and simulations within the CYGNUS collaboration~\cite{Vahsen:2020pzb} showing that a TPC with 3D readout can reach a 10$^5$ gamma/year rejection factor at O(keV). 

While the use of lead can significantly reduce the overall setup dimensions, the~simulation showed that this configuration would require archaeological lead in order not to induce additional background from the shielding, therefore largely raising the cost of this layer. A~cost-benefit optimisation of the shielding materials and thicknesses was hence developed, identifying 2 m of water + 5 cm of copper as the optimal configuration. This shielding provides an attenuation by a factor of 10$^{-7}$ for external photons and 5 $\times$ 10$^{-5}$ for external neutrons, reducing the number of expected electron recoils in the active volume below 10$^3$ cpy (with O(1) cpy nuclear recoils) in the range 1-20~keV. 

For the evaluation of the backgrounds generated by detector components, the~natural photon radioactivity of the parts expected to give the largest contributions was measured with high-purity Germanium detectors thanks to the support of LNGS Special Techniques Service. Results are reported in Table~\ref{tab:camera_radio}. 

\begin{table}[H]
\caption{Measured activity of the internal detector components expected to produce the largest backgrounds in the active volume. The~isotopes in parentheses indicate the activity from that particular part of the decay chain. Upper limits are given at the 90\% confidence~level.}
    \label{tab:camera_radio}
    \resizebox{\columnwidth}{!}{
    \begin{tabular}{ccccccc}
    \toprule
    \textbf{Component} & \boldmath{$^{238}$}\textbf{U}\boldmath{ \textbf{(}$^{234m}$\textbf{Pa) }} &\boldmath{ $^{238}$\textbf{U (}$^{226}$\textbf{Ra) }}&\boldmath{ $^{235}$\textbf{U} } & \boldmath{$^{232}$\textbf{Th (}$^{228}$\textbf{Ra)}} & \boldmath{$^{232}$\textbf{Th (}$^{228}$\textbf{Th) }}&  \boldmath{$^{40}$\textbf{K}}  \\ 
 
\midrule
Camera body [Bq/pc] & 7 & 1.8 & 0.4 & 2.1 & 2.1 & 1.9 \\ 
Camera lens [Bq/pc] & 0.9 & 0.41 & 0.031 & 0.08 & 0.08 & 11 \\ 
GEM foil [Bq/$m^2$] & <0.104 & 0.004 & <0.002 & <0.004 & <0.002 & <0.045 \\ 
Acrylic [Bq/kg] &  &  0.003 &  & 0.005 & 0.004 & 0.035 \\ 
  
 \bottomrule
    \end{tabular}
    }
    
\end{table}

Regarding the GEMs (made of kapton foils, copper clad on each side), the~major source of background is found to come from the frames rather than the foils themselves. For~this reason, in~the LIME prototype of PHASE\_0, the triple $33 \times 33$ cm$^2$ GEMs were mounted and stretched on low-radioactivity acrylic frames, with~the same technique foreseen to be applied to the PHASE\_1~detector. 

For what concerns the sCMOS optical system, a~large $^{40}$K contamination was observed in the lens glass. A~high-purity synthetic fused silica (Suprasil\textsuperscript{\textregistered}) was selected as an alternative material for the fabrication of the lens, for~an expected $\sim$10$^4$ reduction of the contribution from this item. An~overall activity of less than 50 mBq/kg was found in recent measurements performed on a sample at LNGS, confirming the very good properties of this~material.

sCMOS cameras have not been employed yet in DM searches, and~therefore their intrinsic radioactivity has never been studied or optimised in this context. For~this reason, an~extensive program working in close contact with sCMOS camera producer companies and with LNGS Services to assess this aspect within our experimental approach has started. Gamma spectroscopy of several sCMOS cameras as a whole was performed, including models from companies other than Hamamatsu, and~verified that all display similar activities in the O(10) Bq/piece. The~measured activities of the Hamamatsu Orca Fusion are shown in Table~\ref{tab:camera_radio}. A~camera was disassembled in 20 different pieces, which are currently under measurement in order to pinpoint the components introducing the largest radioactivity contamination and possibly replace them by cleaner options. Given the very large sCMOS activity, in~the PHASE\_1 design, they are foreseen to be shielded by the 5 cm copper layer on all sides, except~for the one facing the~GEMs.

Starting from these considerations, a~background evaluation for a 1 m$^3$ PHASE\_1 detector was developed that includes the external gamma and neutron flux contribution with the copper and water shielding discussed above and~the radioactivity contribution of the main internal components. For~the latter,  the~values measured at LNGS for the GEMs, camera lens, camera body and acrylic for the gas vessel and~data from the literature for the Cu of the cathode and the field cage rings~\cite{bib:trex}  were employed. This study showed that O(10$^3$) nuclear recoils and O($10^6$) electron recoils per year are expected in the sensitive volume, with~energy in the 0-20~keV range. It must be noticed that the alphas produced by the GEMs represent the largest contribution to the total NR rate and are absorbed in the gas within 5~cm from them. As~described in Section~\ref{sec:track}, by~exploiting the effect of diffusion in gas, the~track absolute distance from the GEM can be evaluated with a resolution better than 20\%. Therefore, this NR background is expected to be reduced to zero with a suitable selection on the drift distance without~a significant reduction of the detector fiducial volume. Apart from the current lens, the~largest amount of electron recoils is produced by the sCMOS cameras, and, at~a second order, by~the~GEMs.

This study represents the current evaluation of the expected backgrounds for a 1 m$^3$ detector. Therefore it is the starting point for the optimisation and assessment of the background level, which will be further minimised by the detailed study of the GEM, sCMOS and optics~materials.

\subsection{CYGNO PHASE\_2}\label{sec:phase2}

A CYGNO detector with a volume of the order of several tens of cubic meters would be able to generate a significant contribution to the search and study of DM in the mass region below 10 GeV, both for SI and SD coupling. In~case of appearance of signals or evidence of interactions not due to ordinary matter, the~information provided by a directional detector (interaction position, incoming particle direction and energy) will be fundamental to positively confirm the galactic origin of the detected signal as DM and determine its~properties. 

Such a detector could furthermore provide the first directional measurement of solar neutrinos from the pp chain, possibly extending to lower energies the Borexino measurement, as will be illustrated in Section~\ref{sec:neutrino}. 

\begin{figure}[H]
 
 \includegraphics[width=0.7\textwidth]{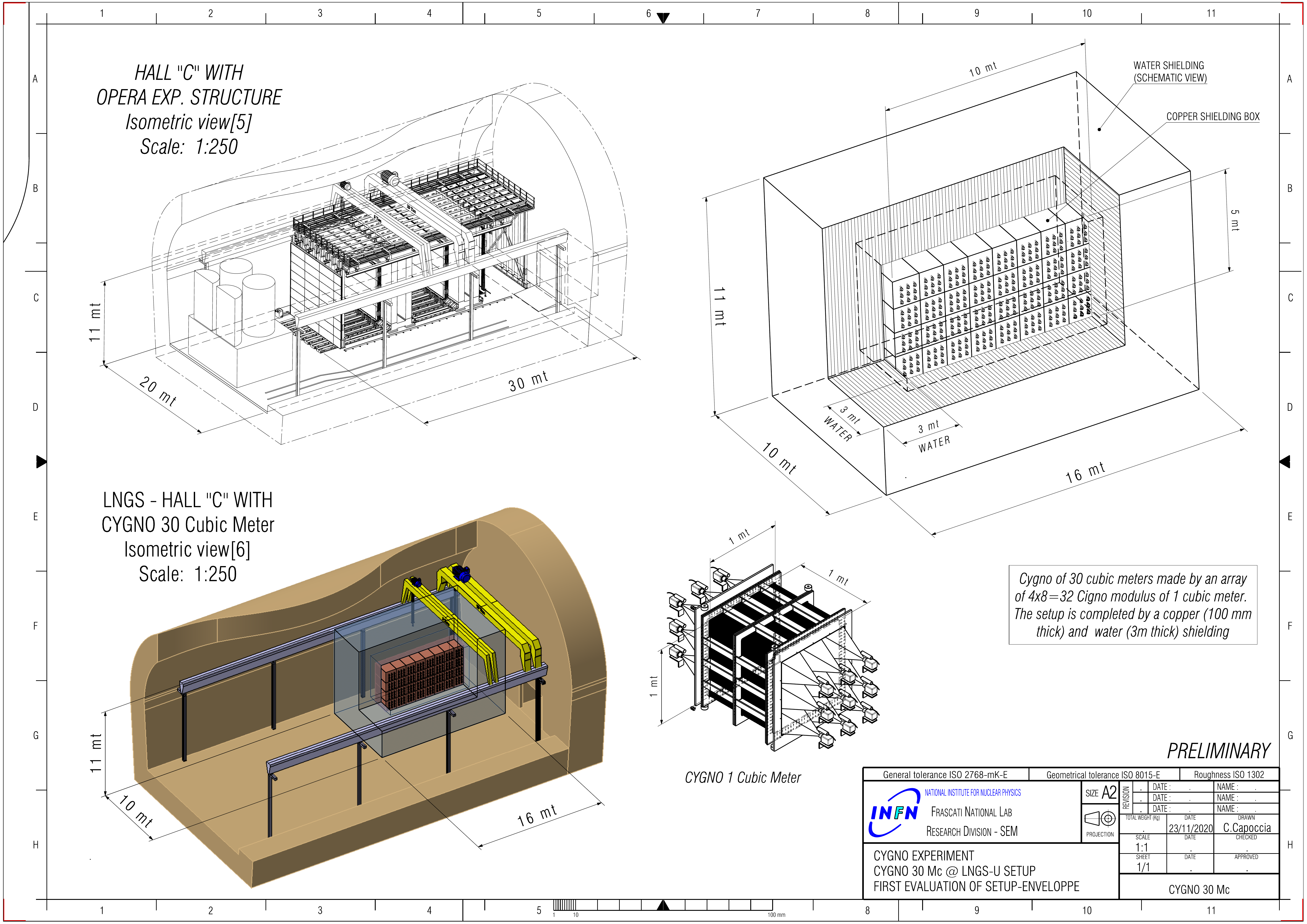}
 \caption{CYGNO PHASE\_2 possible~setup.}
 \label{fig:phase_2}
\end{figure}

As an example, Figure~\ref{fig:phase_2} shows how a 30~m$^3$ sensitive volume apparatus with all its shields (for a total volume of 2$\times 10^3$~m$^3$) would fit into the LNGS experimental Hall~C.

The  development of such a detector in terms of intrinsic background minimisation, performance and costs would, of course, require an improved scalable design, materials and readout. The~possible improvements include, but~are not limited to, the~following:

\begin{itemize}

\item The development of custom sCMOS sensors, with~features focused on CYGNO requirements: low noise, high sensitivity and reduced intrinsic radioactivity together with a lower production cost;
\item The design and realisation of low-radioactivity lenses with fixed focus and large aperture;
\item The reduction of the intrinsic detector material radioactivity, with~the lesson learned after the results obtained with PHASE\_1;
\item The development of innovative gas mixtures for optical readout (illustrated in the following sub-sections) to boost the tracking performances and improve sensitivity for low energy releases. 
\end{itemize}

\subsection{Hydrogen Rich Gas~Mixtures}\label{sec:hydrogen}

The presence of low-mass nuclei as targets in the gas mixture improves detector performance mostly in the low DM mass~region:
\begin{itemize}
    \item Momentum transfer is more efficient, as shown in Equation~(\ref{eq:eps});
    \item Longer lengths of light nuclear recoils in gas produces tracks which are more easy to detect and with a clearer direction.
\end{itemize}

The collaboration is studying the effect of the addition of a small percentage of iC$_4$H$_{10}$ (1--5\%) to the He/CF$_4$ gas mixture. First results demonstrate that, even if the number of photons collected per keV released decreases up to a factor of 3 (with a 5\% addition), light signals are still clear and well-visible.
Studies on this and other hydrogen rich mixtures are still going-on and represent a very promising opportunity to lower the effective DM mass~threshold.

\subsection{INITIUM: an Innovative Negative Ion Time Projection Chamber for Underground Dark Matter~Searches}\label{sec:initium}

The challenging goal of INITIUM is to develop negative ion drift (NID) operation within the CYGNO optical~approach. 

Negative ion drift is a peculiar modification of conventional TPCs (NI-TPC) that involves the addition to the gas of a highly electronegative dopant~\cite{Martoff:2000wi, Ohnuki:2000ex}. In~this configuration, primary electrons produced by an ionizing particle along its track in the gas are captured at very short distances <10-100 $\upmu$m by electronegative molecules, creating negative ions. These anions drift to the anode, where their additional electron is stripped and gives rise to a standard electron avalanche. 
Since anions' mobility depends on mass, the~difference in the time of arrival of different anions effectively provides a measurement of the position of the event along the drift direction. Full 3D detector fiducialization can be obtained by exploiting this information (as achieved in the DRIFT experiment) and~background-free operation over 1 m$^3$ \cite{bib:drift2}. Thanks to these two features, NI-TPC readout planes can image a larger volume than conventional TPC approaches, resulting in lower backgrounds and costs per unit~mass.

The SF$_6$ compound has been recently demonstrated to work very well as a negative ion gas between 20 and 100 Torr, including the possibility of high gains and fiducialization via minority charge carriers~\cite{Phan:2016veo, Ikeda:2020pex, Lightfoot:2007zz}. Compared to the high-vapour pressure, low flash point and low explosive mixture in air of the CS$_2$ employed by DRIFT, SF$_6$ has the substantial advantages of much safer handling, combined with easy radon purification and re-circulation, while at the same time increasing the target fluorine mass. The~studies proved, for the first time, the feasibility of NID at nearly atmospheric pressure (0.8 atm) with He/CF$_4$/SF$_6$ at 360/240/10 Torr with triple thin GEMs and charge pixel readout (Timepix)~\cite{Baracchini:2017ysg}.

The goal of INITIUM is to develop a scintillating He/CF$_4$/SF$_6$-based gas mixture at atmospheric pressure with a low content of SF$_6$ for NID with optical readout. If~NID can be achieved within the optical approach, tracking could be improved by the possibility of reconstructing the track shape along the drift direction by sampling the recorded light at a kHz frame rate. At~the moment, such a high rate can be met only by cameras with low resolution and high noise, which are not yet suited for low-energy rare event searches. Nonetheless, given the fast development of the sCMOS technology, progress in short time is possible, which could open the door to this~possibility.


\section{CYGNO Scientific Goals and Expected Physics~Performances}\label{sec:physics}


A discussion about the expected sensitivity of CYGNO PHASE 1 to WIMP searches and the tools  developed to evaluate it (Section~\ref{sec:wimp}) is presented in the next sections together with an overview of the experiment potentialities towards additional directional~searches.


\subsection{WIMP-like DM Searches at Low Masses through Nuclear Recoil~Signature} \label{sec:wimp}

A statistical analysis based on the Bayesian approach to evaluate the sensitivity of the CYGNO PHASE\_2 30 m$^3$ experiment to WIMP searches in the presence of background was~performed. Details on the used method are described in the Appendix.



The CYGNO approach allows the measurement of both the energy and the direction of the track simultaneously, and~both of these will be combined to evaluate the number of detected events for the final analysis. Nonetheless, since the angular distribution discriminating power is significantly stronger than the energy spectrum shape, this sensitivity study focuses only on the former for the sake of simplicity. In~addition, the~background angular distribution can reasonably be assumed to be isotropic in galactic coordinates, while its energy spectrum will highly depend on the exact materials and shielding employed in the experiment; it is therefore difficult to predict with precision at this stage of the project~development.

However, the energy threshold plays an important role in the determination of the signal angular distribution together with the target nuclei. 
Moreover, while the electrons' kinetic energy (\eVee\ ) is very efficiently translated into ionization in gas, an~important part of slow NR kinetic energy (\eVnr\ ) is loss in scatterings with other nuclei. 
The fraction of \eVnr\ effectively producing ionization is usually referred to as the quenching factor (QF).

In this sensitivity study, two energy thresholds were assumed: a conservative 1~\keVee, backed up by the published results~\cite{bib:fe55}, and~a realistic value of 0.5~\keVee, extrapolated from the improved performances obtained with the PHASE\_0 LIME prototype (see Section~\ref{sec:phase0}). In~order to translate this into nuclear recoil energy, a~SRIM simulation was developed to evaluate the QFs for the elements in our gas mixture, including hydrogen, given the discussion in Section~\ref{sec:hydrogen}. The~QFs for H, He, C and F in He/CF$_4$ 60/40 at 1 atm as a function of the nuclear recoil energy E[\keVnr] were evaluated
and found to be in the range 10\%-30\% for \keVee=100 eV and 60\%-90\% for \keVee=100 \keV.
These result in effective energy thresholds of 1.4 (0.8)  \keVr\  for H, 2.1 (1.2) \keVr\ for He, 3.1 (1.8) \keVr\ for C and 3.8 (2.2) \keVr\ for F for a 1 (0.5) \keVr\ energy deposit.

\begin{figure}[H]
  \vspace{-6pt}
\includegraphics[width=0.7\textwidth]{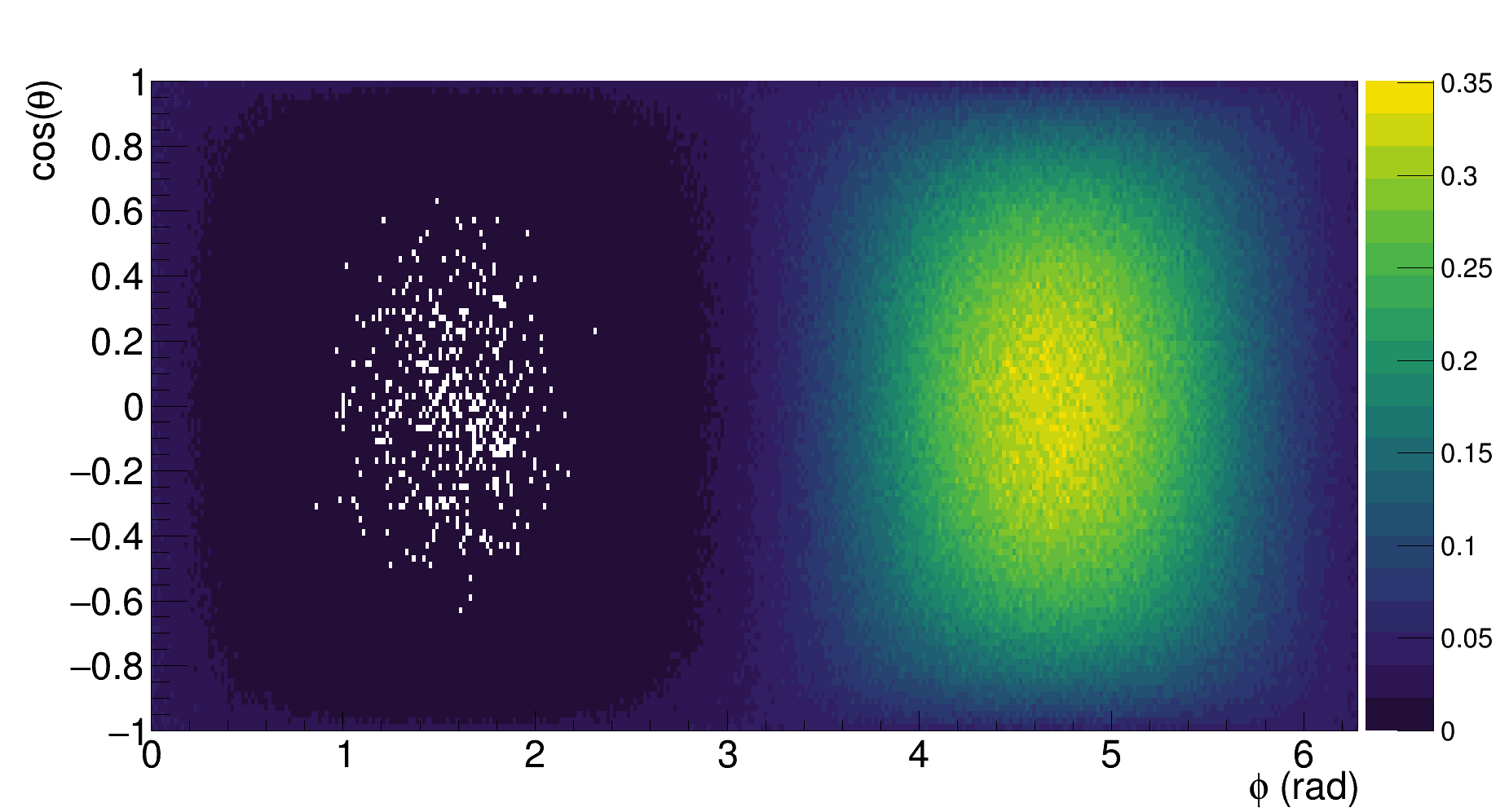} \\
\includegraphics[width=0.7\textwidth]{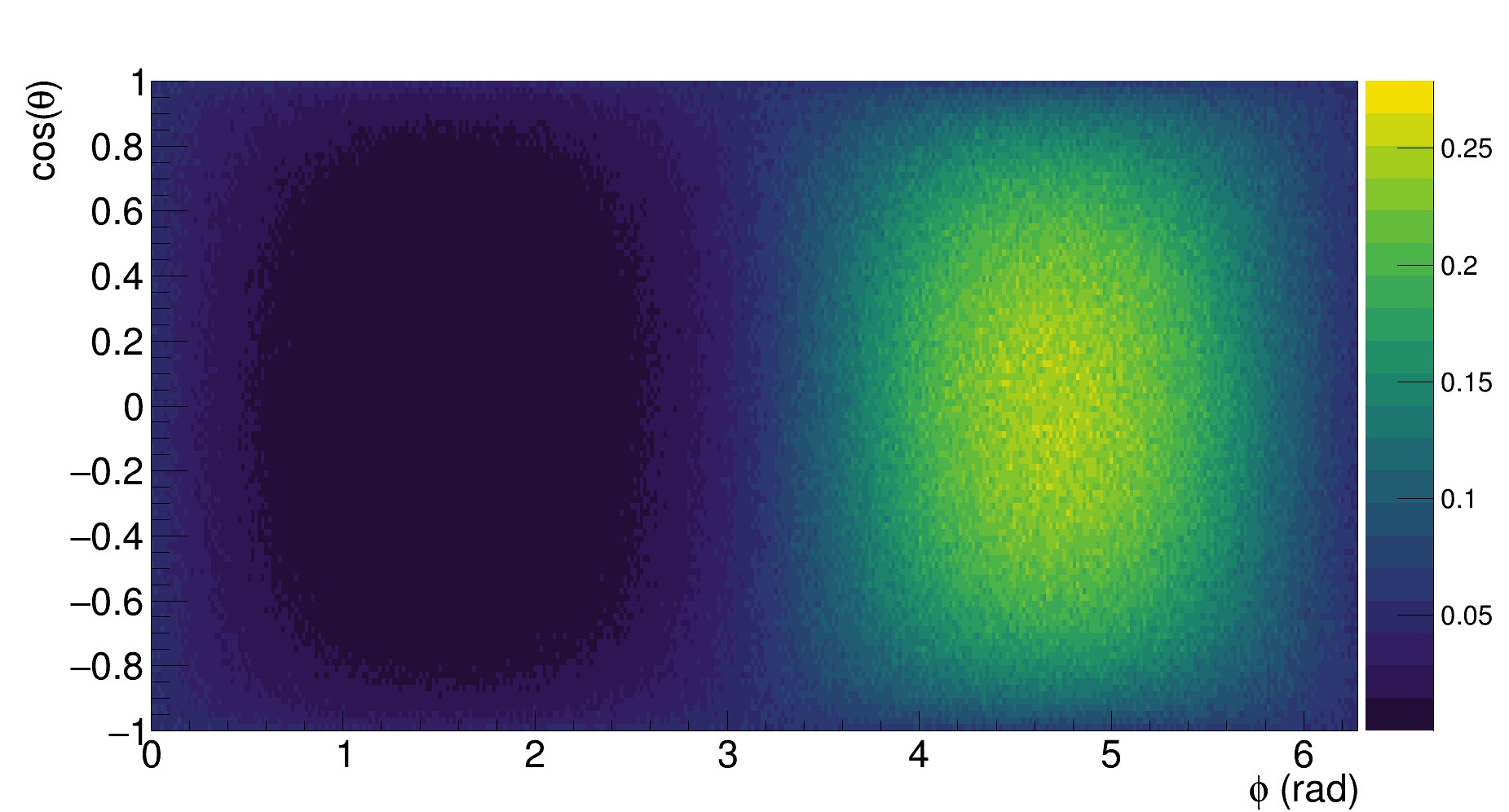}
\caption{Two examples of the angular distribution of recoils due to DM in Galactic coordinates, obtained by Monte Carlo simulations. Top: helium recoils induced by 10 GeV/c$^2$ DM. Bottom: fluorine recoils induced by 100 GeV/c$^2$ DM.}
 \label{fig:spectra2D}
\end{figure}

The signal angular distributions were hence calculated with these effective thresholds in galactic coordinates, starting from~\cite{bib:LEWIN199687,bib:Gondolo_2002,bib:baxter2021recommended} and neglecting the motion of the Earth, as it was shown to have secondary relevance on the angular distribution. Possible shapes of a DM signal nuclear recoil distribution are shown in 2D galactic coordinates in {Figure}~\ref{fig:spectra2D}, 
where the expected anisotropic nature is clearly visible. The~final shape of the distribution strongly depends on three elements: the DM mass, the~element hit and the energy threshold. With~the chosen settings for the analysis, the~angular distributions tend to be strongly peaked at low masses and more spread out at heavier masses, where the angular regions strongly suppressed by kinematics are less~evident.

In order to establish the credible interval (CI) of the sensitivity limits, experiments are simulated by extracting events according to the expected measured angular distributions discussed so far, adding detector effects. Since CYGNO's approach directional capabilities are still under evaluation, for~this study, an angular resolution of $30^\circ$ in the whole detectable range is assumed, as~from literature~\cite{Nakamura:2012zza} and from the CYGNUS simulation~\cite{Vahsen:2020pzb}, with~full head-tail recognition down to the 1 keV$_{ee}$ energy~threshold.

A simple flat distribution of the number of expected signal events between 0 and 1000 is used, given that articulated signal prior probabilities cannot be assumed without risking biases, because~the actual cross-section of DM with protons is unknown. Indeed, events per year is a non-negative defined variable, and due to current limits in the DM community, it is hardly believable that more than 1000 events per year would be produced in the CYGNO~detector.

The number of expected background events for CYGNO PHASE\_2 cannot easily be predicted at this stage of the project, since it will depend on the outcome of PHASE\_0 and PHASE\_1 and the possible improvements discussed in Sections~\ref{sec:phase2}--\ref{sec:initium}. For~this reason, different possible background scenarios are simulated, with~100, 1000 and 10,000 events per year. For~these, a~Poissonian prior is used.  Measurements with PHASE\_1 and simulations informed from these results will, in the future, allow a better estimate of the background~yield.

For each scenario, the~actual number of events is randomly extracted from this Poissonian distribution and a direction is assigned to each, randomly sampling the background angular distribution. After~applying a Gaussian smearing to account for the resolution, a~histogram representing the measured event direction in galactic coordinates is filled, with~its binning reflecting the angular resolution. In~the hypothesis of only background, no events for the WIMP-induced signal recoils are added. In~order to avoid suffering from any underfluctuation of the background (as undersampling), 500 data samples are simulated, and the average result is taken as the final~value. 

The likelihood of the detected events to be the sum of background plus signal (see Section~\ref{sec:appendix} for details) is evaluated on each data sample. From~these, the~posterior probability at 90\%CI of the number of WIMP-induced recoil is computed and~averaged to obtain the final~result. 

\begin{figure}[H]

 \includegraphics[width=0.7\textwidth]{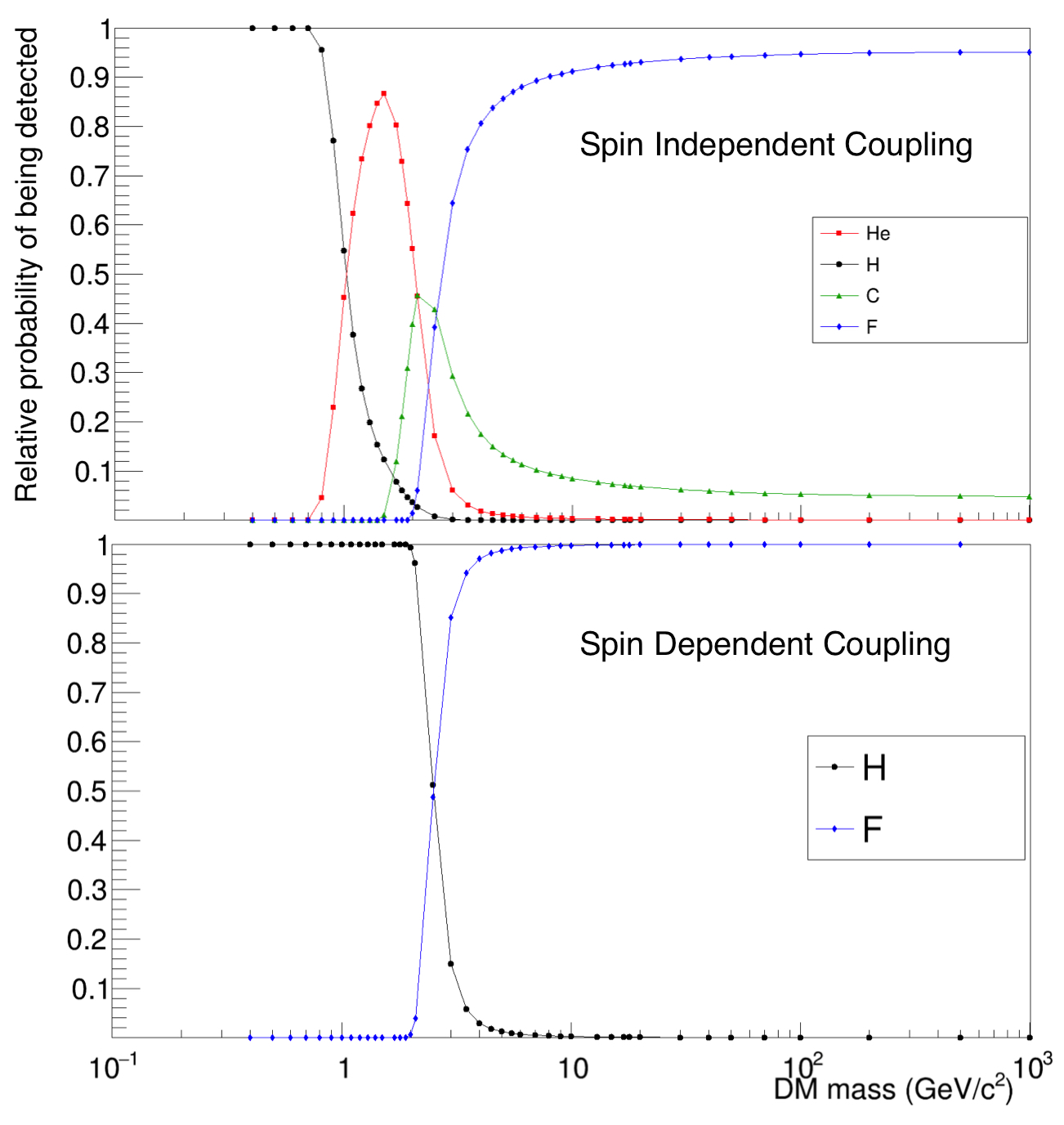} \\

 \caption{Relative probability of nuclear recoils being detected, given that a recoil was detected, as~a function of the DM mass for the SI (\textbf{top}) and SD (\textbf{bottom}) couplings. An~energy threshold of 1 keV$_{ee}$ was used, and the quenching factor corrections are~included.}
 \label{fig:QF_Probele}
 \end{figure}

In order to translate this into a limit in the cross-section versus mass parameters space, it is important to take into account that, because~the target is a mixture of different elements, both the kinematics of the expected DM-nucleus interactions and the expected rate calculation influence the probability of each element to be detected differently as a function of the DM mass. 
This is shown on Figure~\ref{fig:QF_Probele} for a 1 keV$_{ee}$ energy threshold. The~region of the DM velocity distribution accessible to detection is limited at lower values by the energy threshold and at higher values by the local escape velocity (here taken as 544 km/s~\cite{Smith_2007}). For~low DM masses, the~detection of a nuclear recoil is strongly susceptible to the experiment energy threshold. Because~of their light masses, hydrogen and helium detection dominate the early part of the figure, and the rising probabilities of carbon and fluorine reflect their different thresholds. At~higher DM masses, when  the window is quite large, the~A$^2$ cross section enhancement  (where A is the atomic number) dominates, making fluorine the most probable detectable element. Figure~\ref{fig:QF_Probele} also displays the minimum detectable DM mass by each element with a 1 (0.5) keV$_{ee}$ energy threshold; that is, 0.5 (0.3) GeV/c$^2$ for H, 1.0 (0.7) GeV/c$^2$ for He, 1.9 (1.4) GeV/c$^2$ for C and 2.5 (1.9) GeV/c$^2$ for F~recoils.

\begin{figure}[H]

  \includegraphics[width=0.7\textwidth]{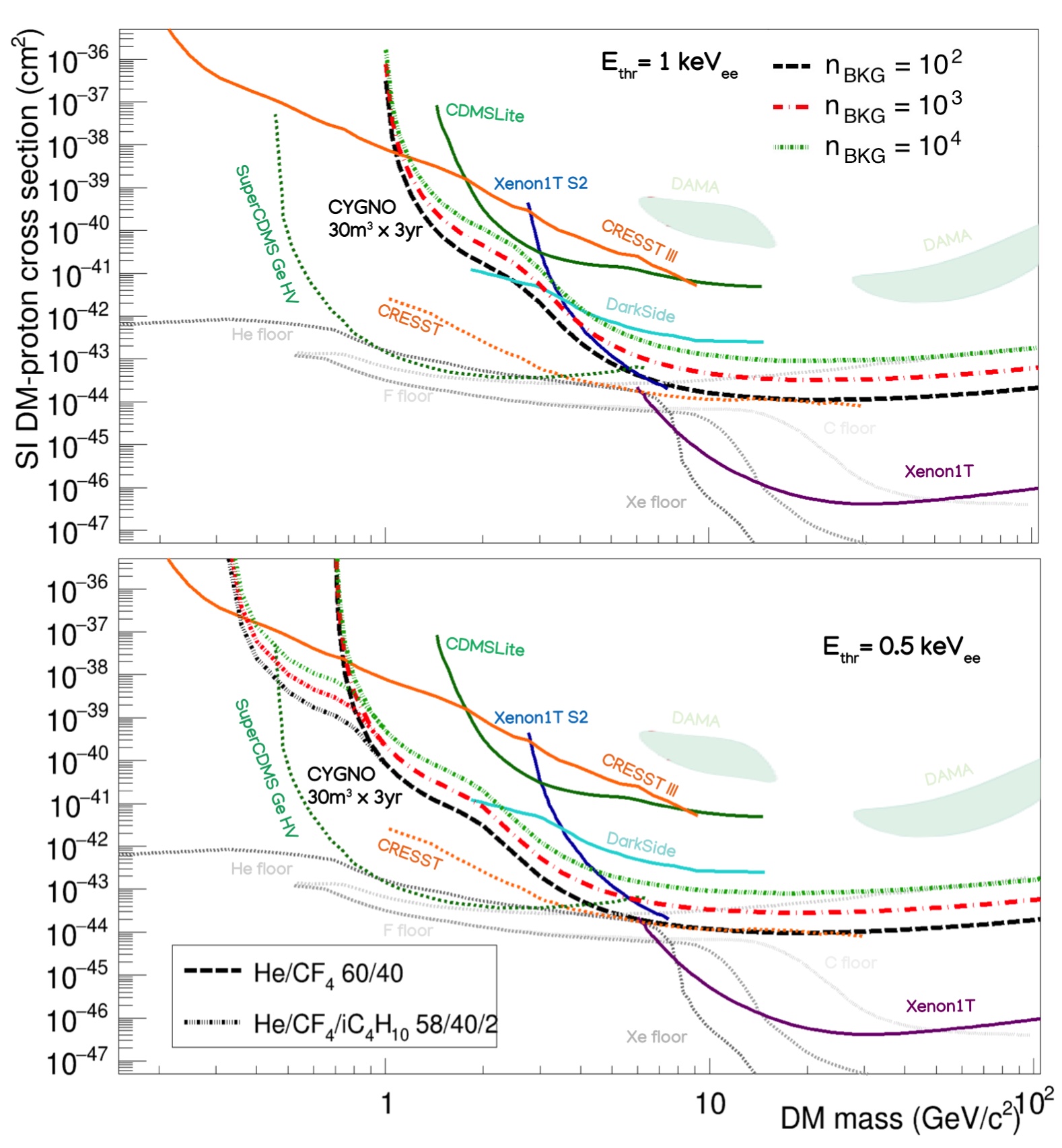} 
 \caption{Spin-independent sensitivity for WIMP-nucleon cross-section for 30 m$^3$ CYGNO detector for 3 years of exposure with different background level assumptions and an operative threshold of 1 keV (top plot) and 0.5 keV (bottom plot). The~dashed curves correspond to a HeCF$_4$ 60/40 detector with N$_{bkg}$ = 100 (black), 1000 (red) and 10,000 (dark green). The~dotted curves show the sensitivity for a HeCF$_4$:isobutane 58/40/2 mixture. Current bounds from Xenon1T (violet) \cite{bib:Aprile_2018}, Xenon1T S2 analysis (blue) \cite{bib:Aprile_2019}, DarkSide (cyan) \cite{bib:Agnes_2018}, CRESST III (orange) \cite{bib:Mancuso:2020gnm} and CDMSLite (green) \cite{bib:Agnese_2018} are also shown. The~densely dotted curves show the future expected limits of SuperCDMS Ge (green)~\cite{SuperCDMS:2016wui} and CRESST (orange) \cite{Willers:2017vae}. The~light gray regions denote DM hints by DAMA~\cite{bib:Savage_2009}, while the different gray curves show the neutrino background levels for different targets~\cite{Boehm:2018sux}.}
 
 \label{fig:SI}
 \end{figure}

Figure~\ref{fig:SI} shows in the top part the expected SI limits for a 30 m$^3$ CYGNO PHASE\_2 experiment for a 3-year exposure with different background scenarios and a 1 keV energy threshold. The possible regions explored with an operating threshold of 0.5 keV are shown in the bottom of Figure~\ref{fig:SI}, together with the results that can be reached with a hydrogen-rich gas mixture with 2$\%$ isobutane content, as~discussed in Section~\ref{sec:hydrogen}.

The shape of the limit reflects the different nuclear composition of the gas mixture. The~lower detectable DM mass obviously corresponds to the one obtainable with the 0.5 keV$_{ee}$ energy threshold and helium quenching factor (hydrogen in the hypothesis of the addition of a small fraction of isobutane). There is a kink on the curve at around 0.9 GeV/c$^2$ corresponding to the transition from hydrogen-dominated to helium-dominated recoils, and~at 3 GeV/c$^2$, from helium- to fluorine-dominated recoils. The~carbon percentage on the total gas mixture (8\%) is too low to produce a visible effect on the~curve.

Figure~\ref{fig:SI} shows how all the scenarios considered in this sensitivity evaluation will be able to probe regions in WIMP masses versus cross-section planes not yet explored, therefore significantly contributing to future DM searches for low WIMP masses. While it is true that the expected reach of future SuperCDMS~\cite{SuperCDMS:2016wui}, CRESST~\cite{Willers:2017vae}, Darkside low-mass~\cite{Darksidelowmass} and~NEWS-G experiments may be able to cover these regions, all of these will be realised through modes of operation that strongly reduce (if not even completely give up) tools for background discrimination. Each of these approaches implies, therefore, very strict (and not yet demonstrated) requirements on the detector materials' radio-purity and the capability to strongly rely on a precise estimate of the expected backgrounds. As~a consequence, any observed signal in this region by these experiments will be difficult to interpret unambiguously as a DM signal. CYGNO's potential of establishing the galactic origin of the detected signal through directional correlation with the Cygnus constellation would therefore constitute a compelling and decisive test to experiment claim in this region, being the only existing approach able to provide a positive identification of a DM signal. CYGNO PHASE\_2 realisation would moreover establish the grounds for the development of a multi-site network of modules for a ton-scale CYGNUS project that, through directionality, could perform a precise study of WIMP properties and DM~astronomy. 

\begin{figure}[H]

 \includegraphics[width=0.7\textwidth]{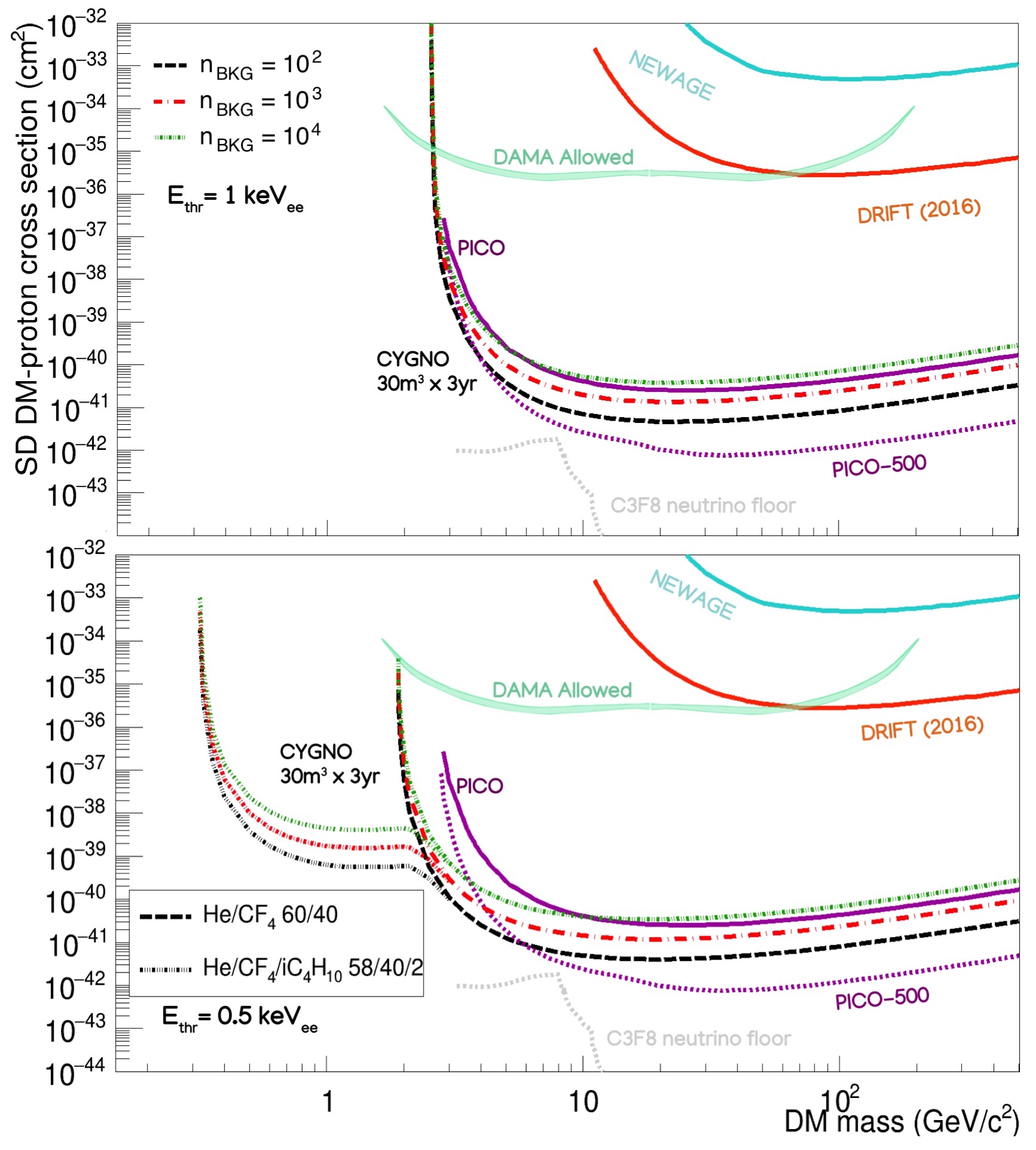} 
\caption{Spin-dependent sensitivity for WIMP--proton cross sections for 30 m$^3$ CYGNO detector for 3 years of exposure with different background level assumptions and an operative threshold of 1 keV (\textbf{top} plot) and 0.5 keV (\textbf{bottom} plot). The~dashed curves correspond to N$_{bkg}$ = 100 (black), 1000 (red) and 10,000 (dark green).  The~dotted curves show the sensitivity for a HeCF$_4$:isobutane 58/40/2 mixture. Current bounds from PICO~(purple) \cite{bib:Amole_2019}, DRIFT (orange) \cite{bib:drift2}, and NEWAGE (cyan) \cite{bib:yakabe2020limits} are also shown. The~allowed region by DAMA is denoted by the light green band~\cite{bib:Savage_2004}. The~light gray dotted line representing the neutrino floor for C$_3$F$_8$ is also taken from PICO \cite{bib:Amole_2019}.} 
 \label{fig:SD}
 \end{figure}

In addition, thanks to the high flourine content, CYGNO PHASE\_2 is expected to also be significantly sensitive to SD couplings and be able to explore regions not yet excluded by the PICO experiment
in the low background scenario, as~shown on the top in Figure~\ref{fig:SD}. 
The PICO experiment, which possesses the strongest sensitivity among all the existing and planned experiments exploring SD coupling, is based, however, on an energy threshold approach. This implies that signal observation does not allow the measurement of the energy of the detected nuclear recoil and could not, therefore, be translated into a constraint in the masses versus coupling parameter space. Hence, also in this context, a~confirmation of the galactic origin of the detected signal would be necessary to establish the properties of the detected WIMP. Moreover, the possibility, now under study, of~running with a threshold of 0.5~keV and adding a small amount (2\%) of isobutane to the gas mixture would allow DM masses even lower than what is expected for the upgrade of PICO to be~reached. 

The estimated sensitivities presented in Figures~\ref{fig:SI} and \ref{fig:SD} demonstrate how CYGNO PHASE\_2 realisation would therefore constitute a very important and compelling step towards the observation and study of a DM signal in the low WIMP mass region for both SI and SD couplings.
\unskip

\subsection{Directional Searches for MeV Dark Matter Produced by Supernovae through Nuclear~Recoil}
While WIMPs still remain highly motivated DM candidates, they are not the only paradigm that can explain the DM presence. Core-collapse supernovae (SN) can reach core temperatures in excess of 30 MeV for O(10) seconds, allowing them to produce vast thermal fluxes of particles with masses O(100) MeV at relativistic speeds~\cite{DeRocco:2019jti}. This makes them an ideal astrophysical source for sub-GeV dark matter. The~DM candidates emerging from this scenario considered in~\cite{DeRocco:2019jti} are dark fermions, but~this is not the only possible realisation of such a mechanism. Those particles end up diffusively trapped near the proto-neutron star that forms from the SN core. The~dark fermions that do eventually escape are produced with a velocity distribution approximately Maxwell--Boltzmann with semirelativistic velocities ({\it v} $\sim$ 1, to~be compared to classical WIMPs with {\it v} $\sim 10^{-3}$  ), exhibiting a roughly order-one spread in velocities. This will result in a time-spreading effect during their propagation to Earth of up to 10$^5$ years for an average galactic SN, creating an overlap in time of various SN. Given the high SN concentration in the galactic center, the~emission of > 100 SN is expected to be overlapping in a diffuse flux at Earth at any given time. This resembles the diffuse flux of SN neutrinos comprising the neutrino floor at energies larger than 10 GeV WIMP~masses.

Thanks to the large dark fermion momentum, such particles, even of masses of O(10 MeV), would cause, in a detector on Earth, a measurable nuclear recoil of O(keV) to be very hard to distinguish from the one induced by a classic WIMP of the galactic halo by an experiment measuring only the energy deposited in the active volume. Nonetheless, the~expected diffuse flux will be strongly peaked towards the galactic center due to the large presence of SN in this region compared to extragalactic sources. Thanks to this high degree of anisotropy, a~directional detector is a crucial tool to discriminate MeV SN-produced DM with respect to classical WIMP scenarios. It has in fact recently been shown that a directional approach with realistic experimental performances could distinguish the two scenarios with few detected signal events, while a non-directional detector typically needs a one to two times order of magnitude more signal yield~\cite{Baracchini:2020owr}. While this study was performed under the assumption of the absence of background in the detected events, a~full estimation of CYGNO sensitivity to this DM candidate scenario with the tools discussed developed for the WIMP physics case in Section~\ref{sec:wimp} is under~development.

\subsection{Solar Neutrino Detection through Both Nuclear and Electron Recoil~Signature}\label{sec:neutrino}
Solar neutrinos are a well-known background to DM searches. They can interact in the active volume of the detector either via elastic scattering on the electrons (producing an electron recoil) or coherent scattering on the nuclei (producing a nuclear recoil). 

Since most current DM experiments possess ER/NR discrimination, typically only the coherent scattering on nuclei is viewed as an irreducible source background, determining the so-called ``neutrino floor''. Directionality has been extensively recognised as the preeminent tool to identify and discriminate NR induced by solar neutrinos from WIMP signal events~\cite{Mayet:2016zxu,Vahsen:2020pzb, Billard:2013qya}. While a ton-scale experiment is needed to start detecting these events~\cite{Vahsen:2020pzb}, due to the low cross-section, new physics in the neutrino sector (described in terms of new mediators between neutrinos and electrons and/or quarks or in terms of non-standard effective interactions) can increase the rate at low energies~\cite{Boehm:2018sux, bib:bertuzzo}. This is particularly true for DM masses below 10 GeV, if~a new scalar mediator is assumed, where the neutrino floor can be raised by several orders of magnitude, making this accessible to CYGNO PHASE\_2. 

Concerning ER induced by neutrino-electron elastic scattering, classical DM experiments measuring only the deposited energy in the detector have no means to discriminate them from ER caused by other sources, and~hence treat them as background. A~detector exhibiting directional capabilities like CYGNO can actually transform these events into a signal. From~the ER direction and the Sun position, the~angle between the incoming neutrino and the scattered electron can in fact be inferred, providing an unambiguous signal identification just like with the one in directional WIMP searches. Ton-scale gaseous TPC have already been proposed in the past~\cite{Seguinot:1992zu,Arpesella:1996uc} to perform solar neutrino spectroscopy. CYGNO's CF$_4$-based gas mixture appears very attractive in this sense~\cite{Arpesella:1996uc} because~it possesses a significative electron density (1.05 $\times$ 10$^{21}$ cm$^{-3}$) with a low $Z$ nuclei, a~feature that maximises the number of targets while minimising multiple~scattering. 

\begin{figure}[H]
 
 \includegraphics[width=0.35\textwidth]{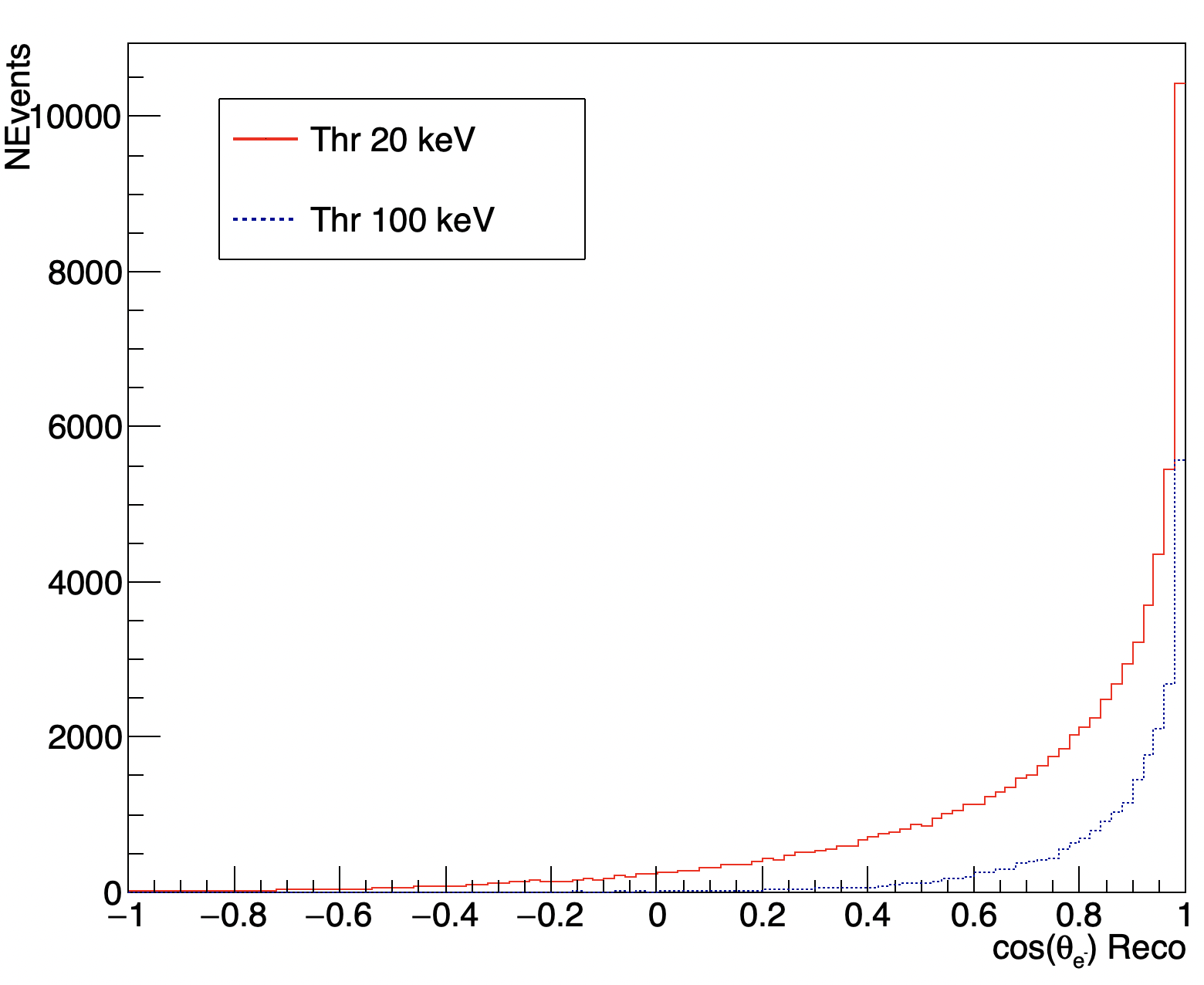}
 \includegraphics[width=0.35\textwidth]{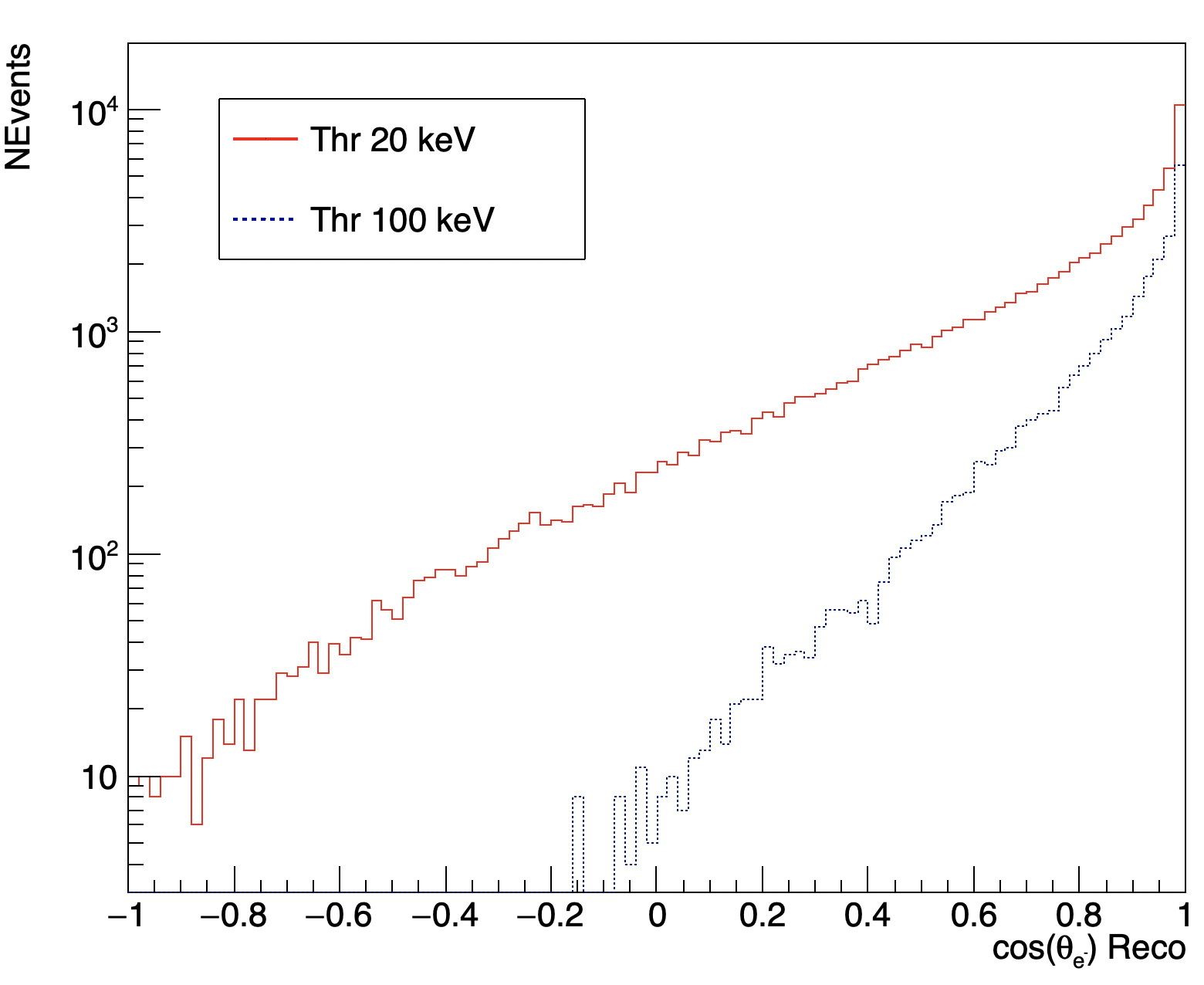}
 \caption{Angular distribution for electron recoils induced by solar neutrinos for 20 keV and 100 keV energy thresholds with a 30$^{\circ}$ $\times$ 30$^{\circ}$ angular resolution, shown on~the right in log~scale.}
 \label{fig:neutrino}
 \end{figure}

About 1 event/m$^3$ per year is expected at atmospheric pressure for an ER energy threshold of 20 keV coming from the $pp$ chain, making this an extremely interesting physics case for the CYGNO PHASE\_2 experiment. Due to the larger multiple scattering suffered by low-energy electrons with respect to nuclei, ER direction determination is more complex than with NR tracks. First results from a dedicated algorithm developed within the collaboration, inspired from X-ray polarimetry~\cite{Soffitta:2012hx}, shows that 30$^{\circ}$ 2D angular resolution with sense recognition larger than 80$\%$ from the sCMOS images analysis can be achieved at 20 keV in a 1 m$^3$ detector, improving at higher energies. Figure~\ref{fig:neutrino} displays the angular distribution for ER induced by Solar neutrinos for 20 keV and 100 keV energy thresholds with a 30$^{\circ}$ $\times$ 30$^{\circ}$ angular resolution. Since background events will be isotropically distributed, this shows how, even with the limited angular resolution assumed, directionality provides an extremely effective means for a high-precision solar neutrino measurement. Moreover, the~20 keV energy threshold assumed for the ER translates to about an 80 keV threshold on the incoming neutrino,
opening a new window of opportunity on the $\emph{pp}$ Sun process down to low energy, unreachable to conventional neutrino detectors~\cite{Seguinot:1992zu}.

\unskip


\section{Conclusions}
In this paper, the~case for directional DM searches with gaseous TPC optically read out through the combination of sCMOS images and PMT signals is presented. The~performances achieved with a 7-litre prototype based on this approach show the possibility of a O(keV$_{nr}$) detection threshold with 10$^2$ ER/NR discrimination at 5.9 keV$_{ee}$.
The CYGNO experiment will develop through a staged approach. The~underground installation at LNGS of a 50~litre prototype (PHASE\_0) is foreseen for the first quarter of 2022, followed by a O(1) m$^3$ experiment (PHASE\_1). PHASE\_0 will allow validation of MC simulation and test CYGNO  performance in an underground environment. PHASE\_1 will be realised to control and minimize the backgrounds from internal materials towards the scalability to larger volume. From~the results of these phases, a~large-scale experiment (30--100 m$^3$) will be proposed to~explore the 1-10 GeV WIMP mass region with high sensitivity to both SI and SD coupling and directionality. A~preliminary sensitivity reach to WIMP searches was hence evaluated for PHASE\_2 with different background assumptions, reflecting realistic performance improvements. This study demonstrates that PHASE\_2 would bring a significant contribution to WIMP searches, not only by probing parameter spaces unexplored so far, but~also being the only approach able to confirm and study any future DM claim by other experiments in this region for~both SI and SD couplings. Additional compelling physics cases which are accessible thanks to directional capabilities have also been discussed, for~which detailed studies are being carried~out.


\funding{This project has received funds under the European
Union’s Horizon 2020 research and innovation programme from the European Research Council
(ERC) grant agreement No 818744 and is supported by the Italian Ministry of Education,
University and Research through the project PRIN: Progetti di Ricerca di Rilevante
Interesse Nazionale “Zero Radioactivity in Future experiment” (Prot. 2017T54J9J).} 


\acknowledgments{The authors want to thank the General Services and Mechanical Workshops of Laboratori Nazionali di
Frascati and Laboratori Nazionali del Gran Sasso  for their precious work and L. Leonzi (LNGS) for technical support. 
}


\appendixstart
\appendix
 \appendixtitles{yes}
\section{Statistical Analysis for the Sensitivity Limit~Evaluation } \label{sec:appendix}

The estimation of the expected limits of the CYGNO experiment was performed applying a Bayesian-based method. In~principle, this approach allows the probability of any model to be calculated given a certain amount of information (data) related to it. This methodology is rarely used in this field, even though it is recently gaining ground~\cite{bib:ROSZKOWSKI200910,bib:TROTTA2007316,bib:Strege_2012,bib:arina2014bayesian,bib:Bringmann_2017,bib:Liem_2016,bib:Messina_2020}.

Every model and parameter which is of interest to the specific analysis, as well as~experimental data, are all considered connected to a probability distribution and, as~such, follow the rules of probability. Exploiting Bayes' theorem, it is possible to find a relation between them and infer a final probability, called posterior, for~the desired quantity. 
In the case of the CYGNO experiment, one is interested in knowing, given a certain number of observed events per year, the~probability that some of them are produced by DM-nucleus interactions.
Those can be events of background ($\mu_b$) or of signal ($\mu_s$), which are strictly connected to the cross-section of WIMP DM particles with protons. Bayes' theorem can be expressed as follows:
\begin{equation}
\label{eq:Bayes}
 p(\mu,\boldsymbol{\theta}\vert \boldsymbol{D},H) = \frac{p(\boldsymbol{D}\vert\mu,\boldsymbol{\theta},H)\pi(\mu,\boldsymbol{\theta}\vert H) }{ \int_{\Omega}\int_{0}^{\infty}p(\boldsymbol{D}\vert \mu,\boldsymbol{\theta},H)\pi(\mu,\boldsymbol{\theta}\vert H)d\mu d\boldsymbol{\theta} } 
\end{equation}
with $p(\boldsymbol{D}\vert\mu_s,\boldsymbol{\theta},H)$ representing the likelihood function $L(\mu_s,\boldsymbol{\theta})$.

In Equation~(\ref{eq:Bayes}), the~following notation is~used:
\begin{itemize}
    \item $p(\mu\vert\boldsymbol{D})$---posterior probability function for the paramenter $\mu$, given $\boldsymbol{D}$;
    \item $\pi(\mu)$---prior probability of a parameter. This includes the expectations of the parameters as well as constraints and knowledge previously obtained from other experiments;
    \item $\mu$---free and of interest parameter representing the expected events due to WIMP-induced recoil ($\mu_s$) or background ($\mu_b$), given a certain WIMP mass (the analysis performs a raster scan);
    \item $\boldsymbol{\theta}$---vector of nuisance parameters, necessary to describe theoretical assumptions and experimental conditions that can affect the results. They can be not completely known and may depend on prior probability distributions. For~example, when $\mu=\mu_s$, $\mu_b$, the~events expected from the background becomes a nuisance parameter;
    \item $\boldsymbol{D}$---data set. Can be made of actual experimental data or simulated data;
    \item $H$---hypothesis under test. It can be the hypothesis of pure background, $H_0$, or~the one where both background and signal are present, $H_1$;
    \item $\Omega$---nuisance parameters space.
\end{itemize}

The likelihood function  used to obtain the results shown in Section~\ref{sec:wimp} is defined as:
\begin{equation}
\label{eq:likelihood}
 p(\boldsymbol{D}|\mu_s,\mu_b,H_1)=(\mu_b+\mu_s)^{N_{evt}}e^{-(\mu_b+\mu_s)  }\prod_{i=1}^{N_{\text{bins}}} \left[ \left( \frac{\mu_b}{\mu_b+\mu_s}P_{i,b}+ \frac{\mu_s}{\mu_b+\mu_s}P_{i,s}\right)^{n_i}\frac{1}{n_i!}\right]
\end{equation}
with:
\begin{itemize}
    \item $N_{evt}$---total number of events of the data sample;
    \item $i$---index representing the bin of the histogram in the 2D angular galactic coordinates;
    \item $n_i$---number of events occurring in the i-th bin;
   \item $\mu$---the~expected events due to WIMP-induced recoil ($\mu_s$) or background ($\mu_b$), given a certain WIMP mass;
  \item $P_{i,x}$---the probability of single event to end up in the i-th bin, according to x model (background or signal).
    
\end{itemize}

The $P_{i,x}$ marginalized probability includes the effects due to the theoretical angular distribution, the~migration from one bin to another caused by resolution effects, and~which element recoils. 
Being in the context of estimating the limits of the CYGNO experiment when data results are consistent with a pure background hypothesis, once the posterior probability of the parameter $\mu_s$ is evaluated, it is possible to compute the upper bound as the 90\% credible interval (C.I.). This is defined as follows:
\begin{equation}
\label{eq:CI}
 \mu_s(90\% CI): \int_{0}^{\mu_s(90\%CI)} p(\mu_s\vert\boldsymbol{D},H_1)d\mu_s=0.9
\end{equation}
where $p(\mu_s\vert\boldsymbol{D},H_1)$ is the posterior probability marginalised over the nuisance parameters. This value represents the limit under which the true value of $\mu_s$ is, with~a 90\% probability.

\reftitle{References}

\end{paracol}
\end{document}